\documentclass[12pt,letterpaper]{article}
\pdfoutput=1
\usepackage[colorlinks=false,
   linkcolor=red, 
   citecolor=blue,
    filecolor=red,
    urlcolor=red,
    linktoc=all, %%%
    pdfstartview=FitV,
    bookmarksopen=true]{hyperref}
\usepackage[left=2cm,top=1cm,right=3cm,nohead]{geometry}

\usepackage[utf8]{inputenc} 
\usepackage{epsfig,latexsym,amsfonts,amsmath,amsthm,amssymb,amsbsy,multirow,slashed,color,mathrsfs,wasysym,textcomp,subfigure,wrapfig,comment,bbold,array,longtable,multirow}

\usepackage{graphicx}
\usepackage{todonotes}
\usepackage{setspace}
\usepackage{subfigure}
\usepackage[boxsize=0.5em,aligntableaux=center]{ytableau}
\usepackage[bf]{caption}
\usepackage{tikz}
\usetikzlibrary{positioning}
\usetikzlibrary{intersections}
\usetikzlibrary{shapes.misc}
\usetikzlibrary{calc}
\newlength{\PicScale}
\usepackage{breqn}
\usepackage[UKenglish]{babel}
\usepackage[toc,page]{appendix}
\usepackage{hhline}
\usepackage{cite}
\usepackage[vcentermath]{youngtab}
\usepackage{datetime}
\usepackage{bbm} 
\hypersetup{
    pdftitle={},
    pdfauthor={},
    pdfsubject={}
} 
\newcommand{\fund}{\mathbf{fund}}

\newcommand{\be}{\begin{equation}}
\newcommand{\ee}{\end{equation}}
\newcommand{\bmat}{\left(\!\!\begin{array}}
\newcommand{\emat}{\end{array}\!\!\right)}

\newcommand{\CC}{{\mathbb{C}}}
\newcommand{\ZZ}{{\mathbb{Z}}}
\newcommand{\PP}{{\mathbb{P}}}
\newcommand{\kod}[1]{\mathrm{#1}}
\newcolumntype{M}[1]{>{\centering\arraybackslash}m{#1}}
\newcolumntype{N}{@{}m{0pt}@{}}

\newcommand{\msm}[1]{\mbox{\small$#1$}}

\numberwithin{equation}{section}

\makeatletter
\def\@cline#1-#2\@nil{%%%
  \omit
  \@multicnt#1%%%
  \advance\@multispan\m@ne
  \ifnum\@multicnt=\@ne\@firstofone{&\omit}\fi
  \@multicnt#2%%%
  \advance\@multicnt-#1%%%
  \advance\@multispan\@ne
  \leaders\hrule\@height\arrayrulewidth\hfill
  \cr
  \noalign{\nobreak\vskip-\arrayrulewidth}}
\makeatother
\graphicspath{{./Images/}}

\hypersetup{
    pdftitle={Non-Geometric Vacua of the Spin(32)/Z2 Heterotic String and Little String Theories},
    pdfauthor={Anamaria Font, Christoph Mayrhofer},
    pdfsubject={non-geometric compactifications; heterotic/F-theory duality; little string theories }
} 
\begin{document}

\thispagestyle{empty}
\vspace*{1.2cm}
\begin{center}
{\LARGE {\bf Non-Geometric Vacua of the $\mathbf{\text{Spin}(32)/\ZZ_2}$ \\[3mm] 
Heterotic String and Little String Theories}}
\vskip1cm 
Anamar\'ia Font$^{a,c}$ and Christoph Mayrhofer$^{b}$\\
\vskip0.8cm
\emph{$^a$ Departamento de F\'isica, Centro de F\'isica Te\'orica y
  Computacional\\
Facultad de Ciencias, Universidad Central de Venezuela\\
A.P. 20513, Caracas 1020-A, Venezuela}
 \vskip0.3cm
\textit{$^b$ Arnold Sommerfeld Center for Theoretical Physics,\\
Theresienstra\ss e 37, 80333 M\"unchen, Germany}\\
\vskip0.3cm
\textit{$^c$Max-Planck-Institut f\"ur Gravitationsphysik, Albert-Einstein-Institut, \\
Am M\"uhlenberg 1, D-14476 Golm, Germany}\\
\vskip0.3cm
\noindent {\small{\texttt{afont@fisica.ciens.ucv.ve, 
christoph.mayrhofer@lmu.de}}}
\vspace{1cm}
\end{center}

\begin{abstract}

\noindent
We study a class of  6d $\mathcal{N}=(1,0)$
non-geometric vacua of the $\mathrm{Spin}(32)/\ZZ_2$ heterotic string
which can be understood as fibrations of
genus-two curves over a complex one-dimensional base.
The 6d $\mathcal{N}=(1,0)$ theories living on the defects that arise when
the genus-two fiber degenerates at a point of the base are analyzed
by dualizing to F-theory on elliptic K3-fibered non-compact
Calabi-Yau threefolds. We consider all possible degenerations of genus-two
curves and systematically attempt to resolve the singularities of the dual 
threefolds. As in the analogous non-geometric vacua of the $E_8\times E_8$ heterotic string,
we find that many of the resulting dual threefolds contain singularities which do not admit a crepant resolution. 
When the singularities can be resolved crepantly, we determine the emerging effective theories which turn out
to be little string theories at a generic point on their tensor branch.
We also observe a form of duality in which
theories living on distinct defects are the same. 

\end{abstract}

\newpage

\tableofcontents

%%%

\section{Introduction}
\label{sec:intro}

The construction and understanding of string compactifications beyond the supergravity approximation are important open problems that 
deserve investigation. For one reason, non-geometric string vacua exist and are in principle on the same footing as the more widely
explored geometric vacua that can be interpreted in terms of supergravity reductions on some internal spaces.
Moreover, it is conceivable that they could have appealing phenomenological features. 
In fact, in the article that originally contemplated the class of non-geometric vacua that we will consider, one motivation 
was to search for compactifications with a reduced number of massless moduli in the low-energy theory \cite{Hellerman:2002ax}.

In \cite{Hellerman:2002ax} the key idea was to build vacua as fibrations by letting the moduli of type II strings compactified on $T^2$ 
vary over a base. A further essential ingredient was to allow for monodromies in the duality group when going around
points on the base where the moduli become singular. Since among these monodromies there are transformations that 
invert the torus volume, the compactifications are intrinsically non-geometric.
The scheme of fibrations of $T^2$ moduli was later extended to heterotic strings where duality with
F-theory can be used to extract properties of the resulting non-geometric vacua \cite{McOrist:2010jw}. 
Taking the base to be complex one-dimensional leads to six-dimensional non-geometric heterotic vacua
that have received attention more recently
\cite{Malmendier:2014uka,Gu:2014ova, Font:2016odl, Malmendier:2016hji, Garcia-Etxebarria:2016ibz}. 

In this paper, we will further examine six-dimensional $\mathcal{N}=(1,0)$ non-geometric heterotic vacua described locally as $T^2$ fibrations over a base ${\mathcal B}$ parametrized by $t \in \CC$. 
As in recent works, we focus on configurations in which the heterotic gauge background is chosen to have $SU(2)$ 
structure so that the gauge group is broken to $E_7 \times E_8$ or $\mathrm{Spin}(28)\times SU(2)/\ZZ_2$, depending on 
whether one starts with the $E_8 \times E_8$ or the $\mathrm{Spin}(32)/\ZZ_2$ heterotic string.
In this situation the heterotic moduli comprise one complex Wilson line together with the complex structure and the
complexified K\"ahler modulus of $T^2$. The T-duality group acting on the space spanned by these three moduli
is $O(2,3,\ZZ)$ \cite{Narain:1985jj}. Restricting this group to $SO^+(2,3,\ZZ)$, the subgroup of order four which can be identified with $Sp(4,\ZZ)$, gives an isomorphism between the heterotic moduli space and the moduli space 
of genus-two curves \cite{Malmendier:2014uka,Gu:2014ova}.
Hence, the non-geometric heterotic vacua can be  defined equivalently as fibrations of a genus-two Riemann surface over the base
${\mathcal B}$. In general, a non-trivial holomorphic fibration will degenerate at certain points on ${\mathcal B}$ and encircling them
will induce an $Sp(4,\ZZ)$ T-duality transformation on the moduli, thereby signalling the presence of defects, dubbed
T-fects in \cite{Lust:2015yia}, such as NS5s or more exotic 5-branes. 
Now, the possible degenerations of genus-two fibers over the \mbox{$t$-plane} have been classified by Ogg, 
and Namikawa and Ueno \cite{ogg66,Namikawa:1973yq}. 
Our objective is to continue the study, initiated in \cite{Font:2016odl}, of the six-dimensional theories living on the
T-fects corresponding to degenerations in the Namikawa-Ueno list.
This program is carried out by dualizing the configuration to F-theory, where the T-fects can be characterized 
geometrically. 

The fundamental heterotic/F-theory duality relates the heterotic string compactified on $T^2$ and F-theory
compactified on an elliptically fibered K3 surface \cite{Vafa:1996xn, Morrison:1996na, Morrison:1996pp}.
At the moduli level, the explicit map when there are no Wilson lines, i.e.\ the gauge group on the heterotic side is unbroken,
was found in \cite{LopesCardoso:1996hq,McOrist:2010jw}.
In the case at hand, when there is one Wilson line breaking the gauge group to  
$E_7 \times E_8$ or $\mathrm{Spin}(28)\times SU(2)/\ZZ_2$, 
the map from the heterotic moduli to the dual K3 moduli was established lately in
\cite{Clingher:2146c,Clingher:3503c,Malmendier:2014uka,Gu:2014ova}.
This map can be expressed in terms of Siegel modular forms of the genus-two curve encoding the heterotic moduli.
Thus, in F-theory the non-geometric heterotic vacua described as genus-two fibrations over a base correspond  to
specific K3 fibrations over the same base. Moreover, to preserve supersymmetry the total space of the F-theory fibration 
must be a Calabi-Yau---a threefold in the case that the base of the K3-fibration is complex one-dimensional.
Since in F-theory there is a well-defined geometric formalism to analyze 
degenerations of the fiber along the base, the heterotic/F-theory duality enables us to infer properties of the T-fects
of non-geometric heterotic vacua.

The \mbox{T-fects} connected to the genus-two degenerations in the Namikawa-Ueno list \cite{Namikawa:1973yq}
were surveyed in \cite{Font:2016odl} in the context of the $E_8\times E_8$ heterotic string. The purpose of the present work is to extend the analysis
to the $\mathrm{Spin}(32)/\ZZ_2$ heterotic string. One motivation is to check for the existence of dualities among defects
observed in \cite{Font:2016odl}. The ultimate goal is to discover the main features of the theories living on the T-fects.
The study of such theories in the $\mathrm{Spin}(32)/\ZZ_2$ heterotic string actually started with the seminal treatise
of Witten \cite{Witten:1995gx} who showed that a heterotic 5-brane, or equivalently a small instanton, supports a
six-dimensional (1,0) supersymmetric gauge theory with group $Sp(1)$ and 16 hypermultiplets in the fundamental representation. 
Already exploiting tools of F-theory, the theories arising from 
$\mathrm{Spin}(32)/\ZZ_2$ small instantons sitting at ADE singularities in K3 were later analyzed  in great detail in
\cite{Aspinwall:1996nk, Aspinwall:1996vc, Aspinwall:1997ye}.
These generic theories were also derived from the dual perspectives of type I D5-branes \cite{Intriligator:1997kq, Blum:1997mm},
and type IIA configurations of D6, D8 and NS5-branes \cite{Brunner:1997gk, Hanany:1997gh}. 
In the $\mathrm{Spin}(32)/\ZZ_2$ heterotic string, F-theory methods were also used early on in \cite{Candelas:1997pq}. 
Various aspects of non-geometric vacua of the $\mathrm{Spin}(32)/\ZZ_2$ heterotic string have been considered more recently in
\cite{McOrist:2010jw, Malmendier:2014uka, Malmendier:2016hji, Garcia-Etxebarria:2016ibz}.

In the following, we will present the results of a systematic study of $\mathrm{Spin}(32)/\ZZ_2$ heterotic \mbox{T-fects} 
associated to the genus-two degenerations in the Namikawa-Ueno classification \cite{Namikawa:1973yq}. 
As in the $E_8 \times E_8$ case addressed in \cite{Font:2016odl}, we will apply the duality map 
to every genus-two degeneration in the Namikawa-Ueno list in order to obtain the dual
F-theory background. Since this background turns out to have a elliptic fibration with a non-minimal singularity, we will attempt to turn the singularity into a minimal one 
by performing a series of blow-ups in the base of the fibration. 
When the resolution can be accomplished we will determine the emerging smooth geometry. 
Introducing blow-ups is equivalent to giving generic vevs to scalars in tensor multiplets of the 6d $\mathcal{N}=(1,0)$ theory on the defect, i.e.\ we move onto the tensor branch of the theory. Thus, knowing
the smooth geometry allows to deduce the gauge groups and matter content characterizing the IR limit, valid in the tensor branch,
of the theory living on the defect. Analogous %%%
techniques have actually been employed in the recent classification 
of SCFTs \cite{Morrison:2012np, Morrison:2012js, Heckman:2013pva,DelZotto:2014hpa,Heckman:2015bfa}
and little string theories (LSTs) \cite{Bhardwaj:2015oru}. 
Actually, the theories that we obtain fall into known configurations whose UV completions are conjectured to be LSTs
 \cite{Bhardwaj:2015xxa, Bhardwaj:2015oru}. The theories indeed have a mass scale and enjoy T-duality upon circle compactification, 
 both typical properties of LSTs \cite{Seiberg:1997zk}. 

Let us finally give an overview of the rest of the article.
The six-dimensional non-geometric heterotic vacua of interest are described in more detail in
Section~\ref{sec:basics}. There we recall the basics of heterotic compactifications on $T^2$ and
review the formulation of heterotic/F-theory duality in terms of a map
between genus-two (sextic) curves and elliptically fibered K3 surfaces.  In addition, we explain how the map connects degenerations
of sextics over a complex one-dimensional base, classified by Namikawa and Ueno, with degenerations of
K3 fibered Calabi-Yau threefolds. 
In Section~\ref{sec:resolution} we first sketch the procedure to resolve singularities
and then apply the method to local heterotic degenerations which
have a geometric description in some duality frame. We also discuss truly non-geometric 
singularities that exhibit a kind of duality with geometric defects.
In Section~\ref{sec:catalog} we catalog all possible local heterotic degenerations admitting 
F-theory duals that can be resolved into smooth Calabi-Yau threefolds. 
We conclude with further observations about the results. Appendix~\ref{app:ADE} contains the
resolutions of several models corresponding to small instantons on ADE singularities.

\section{Non-Geometric Heterotic Vacua and F-Theory}
\label{sec:basics}

This section is devoted to outlining the construction of the six-dimensional non-geometric heterotic vacua studied
in this paper. We will first explain the structure of the vacua and then discuss how to exploit F-theory/heterotic duality
to analyze their properties.

\subsection{Heterotic Vacua in 8 and 6 Dimensions}

The starting point is the compactification of the heterotic string on a torus $T^2$. The emerging eight-dimensional theory
contains moduli fields encoding the geometric and gauge bundle data. The geometric moduli consist of the
complexified K\"ahler modulus, $\rho=\int_{T^2} B+\omega\wedge\bar\omega$,
 with $B$ the Kalb-Ramond two-form and $\omega$ the holomorphic one-form of the torus which follows
 from the metric on $T^2$, and 
the complex structure modulus given by $\tau=\int_b \omega/\int_a \omega$, where $a$ and $b$ are the two 
 generators of the non-trivial one-cycles of the torus.
Furthermore, from the gauge bundle data we have 16 complex Wilson line moduli from the Cartan generators of the non-Abelian gauge group of the 
heterotic string, i.e.\ $\beta^I=\int_a A^I +i\int_b A^I$, $I=1,\dots,16$. 
In the following, we restrict ourselves to background gauge bundles which only have $SU(2)$ structure, so it will
break $E_8\times E_8$ down to $E_8\times E_7$, or $\mathrm{Spin}(32)/\ZZ_2$ to $\mathrm{Spin}(28) \times SU(2)/\ZZ_2$. 
With this choice there is only a single complex modulus, called $\beta$ in the following, whose real and
imaginary parts are given by the Wilson line of the $SU(2)$ Cartan
around the one-cycles of the $T^2$ as defined above. 
It is well known that the three complex parameters $\rho$, $\tau$ and $\beta$ live on the heterotic moduli space \cite{Narain:1985jj}
\begin{equation}
\label{eq:mhet}
 {\mathcal M}_{\text{het}}=O(2;{\mathbb R})\times O(3;{\mathbb R})\backslash O(2,3;{\mathbb R})/O(2,3;{\mathbb Z})\, ,
\end{equation}
where $O(2;{\mathbb R})\times O(3;{\mathbb R})\backslash O(2,3;{\mathbb R})$ is the local moduli space of the $T^2$ compactification and $O(2,3;{\mathbb Z})$ the duality group which identifies physically equivalent theories.

Having the eight-dimensional moduli from the torus compactification we construct, in the next step, six-dimensional vacua by letting the moduli fields vary along two real, or one complex, dimension. Since we allow in this construction for identifications of the moduli under the duality around paths of non-trivial homotopy, it is very cumbersome to work directly with the moduli fields. To circumvent this difficulty we use, like in F-theory, a geometric object which has (almost) the same moduli space as the fields we want to describe.
%%%
%%%
The variation of the fields becomes then a fibration of the object along the complex one-dimensional base. In our case, the geometrification is done via a genus-two curve such that we end up with a genus-two fibration.

Since the description of the heterotic moduli space in terms of genus-two curves will be crucial in the following, we briefly review it.
To every point of the moduli space characterized by  $\rho$, $\tau$ and $\beta$, there is an associated genus-two curve $\Sigma$
whose period matrix $\Omega$ belongs to $\mathbb H_2$ defined by
\begin{equation}\label{eq:Siegel-upper-half-plane}
 \mathbb H_2=\left\{\Omega=\left(\begin{array}{cc} \tau &\beta \\\beta & \rho\end{array}\right)\Big| 
 \det{\text {Im}}(\Omega)>0, \, {\text {Im}}(\rho)>0\right\} \, .
\end{equation}
The four independent one-cycles of $\Sigma$ can be chosen to span a canonical homology basis, $a_i$ and $b_j$ with $i,j=1,2$, such that intersection form is symplectic.
The elements of the period matrix are $\Omega_{ij} = \int_{b_i} \omega_j$, where the $\omega_i$ are
holomorphic one-forms of $\Sigma$ with normalization $\int_{a_i} \omega_j = \delta_{ij}$.  
The moduli space is obtained by taking the quotient of ${\mathbb H}_2$ by the $Sp(4,\mathbb Z)$ action
\begin{equation}\label{eq:Sp4Z-trafo}
\Omega\rightarrow (A \Omega +B)(C\Omega +D)^{-1} \qquad \textmd{with}\qquad
 \left(\begin{array}{cc} A & B \\ C & D\end{array}\right)\in Sp(4,\mathbb Z)\,. 
\end{equation}
The  action of $Sp(4,\mathbb Z)$ is induced by changes of the homology basis that preserve the intersection form.
In \cite{Vinberg:2013} it was shown that $Sp(4,\mathbb Z)$ is isomorphic to $SO^+(2,3,\ZZ)$ which is an index four subgroup of the full  Narain duality group $O(2,3,\ZZ)$.\footnote{Note that this also induces a 
four-to-one map between the moduli space of the genus-two curve  $\mathbb H_2/Sp(4,\ZZ)$ and the true heterotic moduli space  
${\mathcal M}_{\text{het}}$. Hence the moduli spaces are only almost identical as mentioned above.} 
%%%

The geometrification of the heterotic moduli in terms of a genus-two curve suggests, as mentioned already, to build 
lower-dimensional vacua by considering  genus-two fibrations. The idea is simply to let the moduli vary adiabatically along a base
${\mathcal B}$. Since we consider six-dimensional vacua, ${\mathcal B}$ has to be complex one-dimensional, locally parametrized 
by a coordinate $t \in \CC$.  For the moduli to fulfill the (BPS) equations of motion 
the fibration must be holomorphic in $t$. Hence, a non-trivial fibration  $\Sigma(t)$ has to degenerate at co-dimension one loci 
on the base. Encircling such a degeneration point, the genus-two fiber returns to itself but
$\Omega(\Sigma(t))$ undergoes an $Sp(4,\ZZ)$ monodromy transformation. Thus, upon transport around a non-contractible loop,
 the heterotic moduli return to their values only up to a duality transformation. Since $Sp(4,\ZZ)$ includes transformations
 such as $\rho \to - 1/\rho$, which exchanges large and small volume, non-geometric vacua are part of these compactifications. A generic
 genus-two degeneration will induce a monodromy involving all three moduli $\rho$, $\tau$ and $\beta$. 
The localized physical objects that lie at the center of the genus-two degenerations are dubbed {\em {T-fects}}, 
which is short for T-duality defects \cite{Lust:2015yia}. 
In the case of the $E_8\times E_8$ heterotic string, the six-dimensional theories that live on T-fects were studied
in \cite{Font:2016odl}. In this paper we extend the analysis to the $\mathrm{Spin}(32)/\ZZ_2$ heterotic string.

Although genus-two fibrations have monodromies only in $Sp(4,\mathbb Z)\subset O(2,3,\ZZ)$, they have the great advantage that their degenerations are classified. 
%%%
This was done more than forty years ago by Ogg \cite{ogg66} and Namikawa and Ueno \cite{Namikawa:1973yq}, who gave a classification of all possible holomorphic degenerations of genus-two fibers over a complex one-dimensional base.  In particular, Namikawa and Ueno (NU) provide explicit local equations of the possible degenerations 
together with the corresponding monodromies. The NU list supplies a large number of T-fects. However, we do not know how to study them directly in terms of the heterotic string. Therefore, we use the duality with F-theory and analyze them in that setting, as we discuss next.

\subsection{F-Theory and Vacua with Varying Moduli}\label{ss:ftheory}

Since the inception of F-theory \cite{Vafa:1996xn}, it has been known that both heterotic strings compactified on $T^2$ 
and F-theory compactified on an elliptically fibered K3 surface are dual  to each other \cite{Morrison:1996na, Morrison:1996pp}.
This duality is best understood in the large volume/stable degeneration limit \cite{Berglund:1998ej,Morrison:1996na}. 
In this limit $\rho\rightarrow i \,\infty$ on the heterotic side, whereas on the F-theory side the K3 degenerates 
into two $\text {dP}_9$ surfaces which intersect each other along a $T^2$. The heterotic modulus $\tau$ 
is the complex structure of the F-theory $T^2$ at the intersection, while the Wilson lines are encoded in the intersection 
points (spectral cover data \cite{Friedman:1997yq}) of the respective nine exceptional curves of the two 
$\text {dP}_9$'s 
with the $T^2$. Such a precise identification with all Wilson line moduli turned on exists so far only in this special limit.
On the other hand,  the duality map is known along the whole moduli space  when there is
none \cite{LopesCardoso:1996hq,McOrist:2010jw} or
only one non-vanishing Wilson line \cite{Clingher:2146c,Clingher:3503c,Malmendier:2014uka,Gu:2014ova}.  
In this article we focus on the latter case where the Wilson line breaks $E_8\times E_8$ down to $E_8\times E_7$, 
or $\mathrm{Spin}(32)/\ZZ_2$ to $\mathrm{Spin}(28) \times SU(2)/\ZZ_2$.

For the duality with the $E_8 \times E_8$ heterotic string the hypersurface describing the elliptically fibered F-theory K3 
takes the form 
\begin{equation}\label{eq:K3-E7xE8}
 y^2=x^3 + (a\, u^4 v^4 + c\,u^3 v^7)\,x\,z^4 + (b\,u^6 v^6 + d\,u^5 v^7 + u^7 v^5)\,z^6 \, ,
\end{equation}
where $x$, $y$, $z$ and $u$, $v$ are the homogeneous coordinates of the fiber ambient variety $\mathbb P_{2,3,1}$ and the base 
$\mathbb P^1$, respectively. 
This K3 has a $\mathrm{II}^*$ singularity at $v=0$ and a $\mathrm{III}^*$ at $u=0$ which correspond to $E_8$ and $E_7$ gauge groups, respectively. 
%%%
%%%

\begin{figure}
\centering
\subfigure[Reflexive section dividing polytope into $E_7$- and $E_8$-top]{
\parbox{.45\textwidth}{\center
\begin{tikzpicture}
%%%
%%%
%%%
 	[x={(0.50035cm, -0.8cm)},
	y={(.5cm, .5cm)},
	z={(0cm, 1cm)},
	scale=1.000000,
	back/.style={loosely dotted, thin},
	edge/.style={color=blue!95!black},
	topedge/.style={color=red, thick},
	facet/.style={fill=blue!95!black,fill opacity=0.800000},
	vertex/.style={inner sep=2pt,circle,draw=green!25!black,fill=green!75!black,thick,anchor=base},
	point/.style={inner sep=2pt,circle,draw=green!25!black,fill=red!75!black,thick,anchor=base}]
%%%
%%%
%%%
%%%
\coordinate (-3.00000, -2.00000, -6.00000) at (-3.00000, -2.00000, -6.00000);
\coordinate (1.00000, 0.00000, 0.00000) at (1.00000, 0.00000, 0.00000);
\coordinate (0.00000, 1.00000, 0.00000) at (0.00000, 1.00000, 0.00000);
\coordinate (-1.00000, 0.00000, 2.00000) at (-1.00000, 0.00000, 2.00000);
\coordinate (-3.00000, -2.00000, 4.00000) at (-3.00000, -2.00000, 4.00000);
%%%
%%%
%%%
%%%
\draw[edge,back] (-3.00000, -2.00000, -6.00000) -- (0.00000, 1.00000, 0.00000);
%%%
%%%
%%%
%%%
%%%
%%%
%%%
%%%
\fill[facet] (-3.00000, -2.00000, 0.00000) -- (1.00000, 0.00000, 0.00000) -- (0.00000, 1.00000, 0.00000) -- cycle {};
%%%
%%%
%%%
%%%
%%%
%%%
\draw[edge] (-3.00000, -2.00000, -6.00000) -- (1.00000, 0.00000, 0.00000);
\draw[edge] (-3.00000, -2.00000, -6.00000) -- (-3.00000, -2.00000, 4.00000);
\draw[edge] (1.00000, 0.00000, 0.00000) -- (0.00000, 1.00000, 0.00000);
\draw[edge] (1.00000, 0.00000, 0.00000) -- (-1.00000, 0.00000, 2.00000);
\draw[edge] (1.00000, 0.00000, 0.00000) -- (-3.00000, -2.00000, 4.00000);
\draw[edge] (0.00000, 1.00000, 0.00000) -- (-1.00000, 0.00000, 2.00000);
\draw[edge] (-1.00000, 0.00000, 2.00000) -- (-3.00000, -2.00000, 4.00000);
%%%
%%%
%%%
%%%
\node[vertex] at (0, 0, 0)     {};
%%%
%%%
%%%
%%%
\node[vertex] at (-3.00000, -2.00000, 0.00000)     {};
\node[vertex] at (1.00000, 0.00000, 0.00000)     {};
\node[vertex] at (0.00000, 1.00000, 0.00000)     {};
%%%
%%%
%%%
%%%
%%%
\draw[topedge] (-1, 0, -2) -- (-3, -2, -6);
\draw[topedge] (-1, -1, -3)  -- (-3, -2, -6);
\draw[topedge] (-3, -2, -1)  -- (-3, -2, -6);
\foreach \z in {-6,-5,...,-1}
     \node[point] at (-3, -2, \z)  {};
\node[point] at (-2, -1, -4)   {};
\node[point] at (-1, 0, -2)    {};
\node[point] at (-1, -1, -3)   {};
%%%
\draw[topedge] (-3, -2, 1)  -- (-3, -2, 4);
\draw[topedge] (-3, -2, 4)  -- (-1, -1, 2);
\draw[topedge] (-3, -2, 4)  -- (-1, 0, 2);
\draw[topedge] (-1, 0, 2) -- (0, 0, 1);
\foreach \z in {1,2,...,4}
     \node[point] at (-3, -2, \z)  {};
\node[point] at (-2, -1, 3)    {};
\node[point] at (-1, 0, 2)     {};
\node[point] at (0, 0, 1)      {};
\node[point] at (-1, -1, 2)    {};
\end{tikzpicture}
}}\hfill
\subfigure[Reflexive section dividing polytope into $SO(28)$- and trivial-top ]{
\parbox{.45\textwidth}{\center
\begin{tikzpicture}
%%%
%%%
%%%
 	[x={(0.50035cm, -0.8cm)},
	y={(.5cm, .5cm)},
	z={(0cm, 1cm)},
	scale=1.000000,
	back/.style={loosely dotted, thin},
	edge/.style={color=blue!95!black},
	topedge/.style={color=red, thick},
	facet/.style={fill=blue!95!black,fill opacity=0.800000},
	vertex/.style={inner sep=2pt,circle,draw=green!25!black,fill=green!75!black,thick,anchor=base},
	point/.style={inner sep=2pt,circle,draw=green!25!black,fill=red!75!black,thick,anchor=base}]
%%%
%%%
%%%
%%%
\coordinate (-3.00000, -2.00000, -6.00000) at (-3.00000, -2.00000, -6.00000);
\coordinate (1.00000, 0.00000, 0.00000) at (1.00000, 0.00000, 0.00000);
\coordinate (0.00000, 1.00000, 0.00000) at (0.00000, 1.00000, 0.00000);
\coordinate (-1.00000, 0.00000, 2.00000) at (-1.00000, 0.00000, 2.00000);
\coordinate (-3.00000, -2.00000, 4.00000) at (-3.00000, -2.00000, 4.00000);
%%%
%%%
%%%
%%%
\draw[edge,back] (-3.00000, -2.00000, -6.00000) -- (0.00000, 1.00000, 0.00000);
%%%
%%%
%%%
%%%
%%%
%%%
%%%
%%%
\fill[facet] (1, 0, 0) -- (-1, 0, 2) -- (-1, 0, -2) -- cycle {};
%%%
%%%
%%%
%%%
%%%
%%%
\draw[edge] (-3.00000, -2.00000, -6.00000) -- (1.00000, 0.00000, 0.00000);
\draw[edge] (-3.00000, -2.00000, -6.00000) -- (-3.00000, -2.00000, 4.00000);
\draw[edge] (1.00000, 0.00000, 0.00000) -- (0.00000, 1.00000, 0.00000);
\draw[edge] (1.00000, 0.00000, 0.00000) -- (-1.00000, 0.00000, 2.00000);
\draw[edge] (1.00000, 0.00000, 0.00000) -- (-3.00000, -2.00000, 4.00000);
\draw[edge] (0.00000, 1.00000, 0.00000) -- (-1.00000, 0.00000, 2.00000);
\draw[edge] (-1.00000, 0.00000, 2.00000) -- (-3.00000, -2.00000, 4.00000);
%%%
%%%
%%%
%%%
\node[vertex] at (0, 0, 0)     {};
%%%
%%%
%%%
%%%
\node[vertex] at (1, 0, 0)      {};
\node[vertex] at (-1, 0, 2)     {};
\node[vertex] at (-1, 0, -2)    {};
\node[vertex] at (0, 0, 1)      {};
%%%
%%%
%%%
%%%
%%%
\draw[topedge] (-2, -1, -4) -- (-3, -2, -6);
\draw[topedge] (-1, -1, -3)  -- (-3, -2, -6);
\draw[topedge] (-3, -2, 4)  -- (-3, -2, -6);
\draw[topedge] (-3, -2, 4)  -- (-2, -1, 3) ;
\draw[topedge] (-3, -2, 4)  -- (-1, -1, 2);
\foreach \z in {-6,-5,...,4}
     \node[point] at (-3, -2, \z)  {};
\node[point] at (-2, -1, -4)   {};
\node[point] at (-1, -1, -3)   {};
\node[point] at (-2, -1, 3)    {};
\node[point] at (-1, -1, 2)    {};
%%%
%%%
\node[point] at (0, 1, 0)    {};
\end{tikzpicture}     
}}
 \caption{The two different reflexive sections through the (dual) polytope defined via equation \eqref{eq:K3-E7xE8} and \eqref{eq:K328a}, respectively. The red dots indicate toric divisors which correspond to the resolution divisors of the $E_7$, $E_8$, and $SO(28)$ gauge groups. Divisors which intersect each other lie next to each other along the red edges. Note the appearance of the extended Dynkin diagram structure. Furthermore, the $SU(2)$ of $\mathrm{Spin}(28)\times SU(2)/\mathbb Z_2$ on the right hand side is given by the divisor corresponding to the green interior point on the edge of the reflexive section.}\label{fig:e7xe8-K3}
\end{figure}
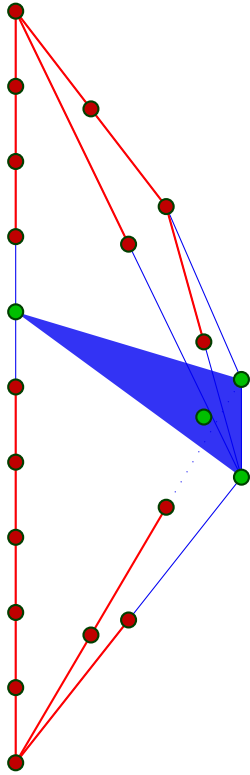
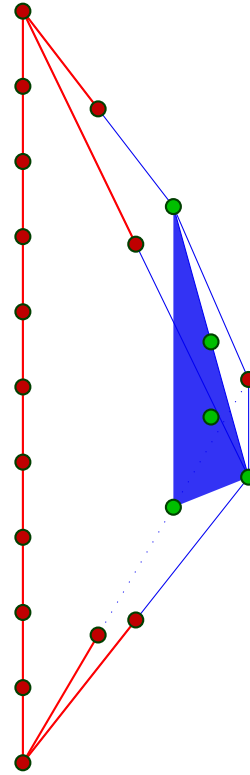

For the $\mathrm{Spin}(32)/\ZZ_2$ heterotic compactification 
the dual K3 is described by  \cite{Aspinwall:1996nk, Aspinwall:1996vc, McOrist:2010jw}
\begin{equation}
y^2=x^3+ Q\,x^2 z^2+\epsilon\, x z^4 \, ,
\label{eq:K328a}
\end{equation}
where $Q$ and $\epsilon$ depend on the base coordinates according to
\begin{equation}
\label{qedef}
\begin{split}
Q & = v( u^3 + a\, u v^2 +b \, v^3) \,,\\
\epsilon & = v^7(c\,u+d\,v)\,.
\end{split}
\end{equation}
The fibrations \eqref{eq:K328a} and \eqref{eq:K3-E7xE8} are birationally equivalent to each other as shown for instance in \cite{McOrist:2010jw, Malmendier:2014uka}, and for an earlier account of this in terms of toric geometry see \cite{Candelas:1997pq}. 
In the latter reference it is described how the two different fibrations are realized as two different two-dimensional reflexive sections of the same polytope. In Figure~\ref{fig:e7xe8-K3} we indicated, for our case, the two different sections of the K3 which give \eqref{eq:K3-E7xE8}  and \eqref{eq:K328a}, respectively. 
Although a given fan of the polytope allows at most for one fibration, they are connected via birational flops.
The double fibration structure is expected from the known T-duality of the two heterotic strings upon circle 
compactification \cite{Aspinwall:1997ye, Bhardwaj:2015oru}.

To analyze the K3 in \eqref{eq:K328a} we bring it to Weierstra{\ss} form. In the patch $z=1$ we obtain
\begin{equation}
y^2=x^3+  \frac13 \left(3\epsilon - Q^2\right) x + \frac{Q}{27}\left(2 Q^2 - 9\epsilon\right) \, .
\label{eq:K328b}
\end{equation}
The singularities of this fibration are located at the vanishing locus of the discriminant
\begin{equation}
\label{discri32}
\begin{split}
\Delta& = -\epsilon^2(Q^2-4\epsilon)\,, \\
&=-v^{16}(cu + dv)^2 \left(u^6 + 2 au^4 v^2 + 2 b u^3 v^3 + a^2 u^2 v^4 + 
(2 ab - 4c) u v^5 + (b^2 - 4 d) v^6\right)\,.
\end{split}
\end{equation}
We observe that the fiber has Kodaira singularities of type $\mathrm{I}^*_{10}$
($SO(28)$) at $v=0$, and of type $\mathrm{I}_2$ ($SU(2)$) at
$cu+dv=0$, for generic coefficients. The gauge group is actually
$\mathrm{Spin}(28)\times SU(2)/\ZZ_2$. When $c\equiv 0$ the group enhances to
$\mathrm{Spin}(32)/\ZZ_2$.

Heterotic/F-theory duality requires the existence of a map relating the heterotic moduli \eqref{eq:Siegel-upper-half-plane} to
the coefficients $a$, $b$, $c$ and $d$ of the dual K3 surfaces \eqref{eq:K3-E7xE8} and \eqref{eq:K328a}.
The duality map has been established recently \cite{Clingher:2146c, Clingher:3503c,Malmendier:2014uka}.
It can be written  as
\begin{equation}
\label{abcd}
a=-\frac1{48}\psi_4(\Omega)\,,\quad b=-\frac1{864}\psi_6(\Omega)\,,\quad c=-4\chi_{10}(\Omega)\,, \quad d=\chi_{12}(\Omega)\,,
\end{equation}
where $\psi_4$, $\psi_6$, $\chi_{10}$, and $\chi_{12}$ are genus-two Siegel modular forms \cite{Bruinier_2008} of 
weight indicated by the subscript.
Modularity is meant with respect to the $Sp(4,\mathbb Z)$ transformation \eqref{eq:Sp4Z-trafo}.

{}From the previous discussion we conclude that there is a well-defined relation between the moduli space of the heterotic
string compactified on $T^2$ with one complex Wilson line, the moduli spaces of genus-two curves and elliptically
fibered K3 surfaces with $E_7 \times E_8$ or $\mathrm{Spin}(28)\times SU(2)/\ZZ_2$ singularities.
Thus, non-geometric heterotic vacua encoded in terms of genus-two fibrations over a base can be realized, in F-theory, as 
K3 fibrations over the same base. The advantage of this correspondence is that in F-theory there is a proper procedure to analyze 
degenerations of the fiber along the base. In this way, heterotic/F-theory duality can be applied to explore the physics
of T-fects associated to genus-two degenerations. 

We are mostly interested in six-dimensional non-geometric heterotic vacua given by genus-two fibrations 
over a base with local coordinate $t$. In particular, we want to consider the genus-two degenerations
compiled by Namikawa and Ueno (NU) \cite{Namikawa:1973yq}. 
In the NU classification the genus-two singularities are described in terms of fibrations of hyperelliptic curves 
represented by sextics of the form
\begin{equation}\label{eq:sextic}
 y^2=c_6(t)\,x^6+c_5(t)\,x^5+\ldots +c_1(t)\,x + c_0 \, ,
\end{equation}
where the $c_i(t)$ are functions (or sections) of $t$. Furthermore, the hyperelliptic curve fibrations of \cite{Namikawa:1973yq} are in a canonical form with
the singularity located at $t=0$. Determining the K3 coefficients $a$, $b$, $c$, $d$, as functions of $t$
is facilitated by having the genus-two fibrations in the form of \eqref{eq:sextic}.
To begin we compute the modular forms of the genus-two curve from the
Igusa-Clebsch invariants\footnote{See appendix C of \cite{Font:2016odl} for the explicit form of the Igusa-Clebsch 
invariants in terms of the coefficients of the sextic.} $I_2$, $I_4$, $I_6$, $I_{10}$ \cite{Igusa:1962}
\begin{equation}
\label{SiegelICmap}
\begin{aligned}
&I_2(c_i) = \frac{\chi_{12}(\Omega)}{\chi_{10}(\Omega)} \,,
&I_4(c_i) = 2^{-4}\cdot 3^{-2} \psi_4(\Omega) \,,\\
&I_6(c_i) = 2^{-6}\cdot 3^{-4} \psi_6(\Omega) + 2^{-4}\cdot 3^{-3}\frac{\psi_4(\Omega) \chi_{12}(\Omega)}{\chi_{10}(\Omega)} \,,
&I_{10}(c_i) = 2^{-1}\cdot 3^{-5} \chi_{10}(\Omega) \,.
\end{aligned}
\end{equation}
Combining these relations with \eqref{abcd} then gives
\begin{equation}
\label{abcdI}
a=- 3I_1, \quad b=2(I_2I_4 - 3 I_6), \quad c=-2^3 3^5 I_{10}, \quad d=-2 \, 3 ^5 I_2 I_{10} \, ,
\end{equation}
for the complex structure coefficients of the K3. Since the Igusa-Clebsch invariants are polynomials of the $c_i(t)$'s 
so will be the coefficients of the dual K3. 
Hence, in the end we obtain for every genus-two singularity in the NU list a K3 fibration over the same $t$-plane with the K3 fiber degenerating at $t=0$ as well. Let us remark here that understanding the map from the F-theory to the heterotic setup is much more 
involved. Some progress in this direction has been achieved recently in \cite{Garcia-Etxebarria:2016ibz}.

In the next section we will look at the F-theory singularities that arise from the map and attempt to resolve them.
It turns out that whether a resolution is possible or not depends on the vanishing order of the coefficients 
$a$, $b$, $c$ and $d$, denoted $\mu(a)$,  $\mu(b)$, $\mu(c)$ and $\mu(d)$, at $t=0$. Notice that in all cases we have  $\mu(c) \le \mu(d)$.

To conclude this section let us briefly consider the fibration of the (by itself elliptically fibered) 
F-dual K3 over a compact base, concretely over a ${\mathbb P}^1$. 
In the case of the $\mathrm{Spin}(32)/\ZZ_2$ heterotic string, cf.~\eqref{eq:K328b}, imposing that the total space is a Calabi-Yau threefold implies that the latter can be understood as an elliptic fibration over the Hirzebruch surface ${\mathbb F}_{4}$.\footnote{Starting with \eqref{eq:K3-E7xE8} in the $E_8\times E_8$ heterotic string leads to an elliptic fibration over ${\mathbb F}_{12}$
\cite{Malmendier:2014uka}.} Moreover, $a$, $b$, $c$, and $d$ must be  
polynomials of degree 8, 12, 20 and 24, respectively, in the homogeneous coordinates of the base. 
Now, it is known that this F-theory compactification is precisely dual to the $\mathrm{Spin}(32)/\ZZ_2$ heterotic string compactified
on K3 with group broken to \mbox{$\mathrm{Spin}(28)\times SU(2)/\ZZ_2$} and 20 half-hypermultiplets in the 
$({\mathbf{28}}, {\mathbf{2}})$ representation \cite{Morrison:1996na, Morrison:1996pp}.
%%%

\section{Resolution of Singularities: Formalism and Examples}\label{sec:resolution}  

After establishing the duality map between heterotic and F-theory vacua the next task is to tackle
the resolution of singularities. To this end we first review a general formalism
based on toric techniques  \cite{Font:2016odl} in this section. Afterwards we apply the method to a class of NU models
whose degenerations correspond to small instantons on ADE singularities. Comparing with
the known resolutions in these cases \cite{Aspinwall:1997ye, Blum:1997mm} allows to verify the validity 
of our approach. We then consider examples in which there is no initial interpretation of the 
singularities.

\subsection{Formalism}\label{ss:toric}

We work systematically with a Weierstra\ss{} model all along. This means that an elliptic fibration is always 
represented by a hypersurface equation of the form
\begin{equation}\label{eq:Weierstrass-eqn}
 y^2=x^3 + f(\xi_i)\,x\,z^4 + g(\xi_i)\,z^6
\end{equation}
where $x$, $y$, $z$ are again the homogeneous coordinates of $\PP_{2,3,1}$ and $f$ and $g$ are sections of some line 
bundles over the base $B\ni \xi_i$, $i=1,2$. The crucial requirement in F-theory is that
the elliptic fibration has to be Calabi-Yau. This condition constrains
the line bundles of $f$ and $g$ to be $K^{-4}_B$ and $K^{-6}_B$, respectively, 
with $K_B$ the canonical bundle of the base. The elliptic fiber becomes singular when the 
discriminant $\Delta=4f^3 + 27 g^2$ vanishes. We will refer to a model as resolved if the elliptic 
curve has only minimal singularities (or Kodaira type singularities) \cite{kodaira123}  along the base, 
i.e.\ there are no non-minimal points along the discriminant locus where $f$ vanishes to order four or higher and 
simultaneously $g$ to order six or higher.  

In the F-theory framework the six-dimensional vacua of interest are obtained by taking a K3
fibration over a base parametrized by $t \in \CC$. Thus, at the start we have a hypersurface equation such as
\eqref{eq:K3-E7xE8} or \eqref{eq:K328b}, with $f$ and $g$ depending on $(u,v,t)$.
Recall that the dependence on $t$ is dictated by the particular NU model under study.

To sketch the resolution procedure let us first examine the F-theory dual of the $E_8 \times E_8$ heterotic string,
which is simpler yet captures the essentials.
Since  the coefficient in front of the $u^7\,v^5$ term in $g$ in \eqref{eq:K3-E7xE8} is constant, there is no non-minimal singularity along $v=0$. Therefore, we only have to look in the $(u,t)$ patch  for such points and in the beginning
it turns out that there is just one non-minimal singularity at $u=t=0$. To get rid of this non-minimal point we follow 
\cite{Aspinwall:1997ye} and blow-up the base at this point. However, as we will explain momentarily, we proceed in a 
rather toric manner by introducing the maximal amount of 
\footnote{Crepant in the sense that the proper transform of  the hypersurface equation 
\eqref{eq:K3-E7xE8}, or \eqref{eq:K328b},  
after the base blow-up is still Calabi-Yau. We do not claim that the canonical class of the base does not change which is obviously wrong.} blow-ups at the non-minimal point at once, and not blow-up after blow-up. 
Subsequently, we search for non-minimal points along the newly introduced exceptional curves and, if necessary, apply 
the toric blow-up method also to these points. The procedure stops when we do not find any new non-minimal points anymore.

We next turn to the dual F-theory K3 of the $\mathrm{Spin}(32)/\ZZ_2$ heterotic string.
In the defining equation \eqref{eq:K328b} the coefficients of $u^9 v^3$  in $g$ and  $u^6 v^2$ in $f$ are both constant so
that non-minimal points along $v=0$ are absent. Thus, again it suffices to work in the $(u,t)$ patch.
A hallmark of this string is that in all NU models there is a singularity along $t=0$. 
Moreover, the vanishing degrees of $(f,g,\Delta)$ and the
monodromy cover along this curve indicate a singular fiber of type $I_{2k}$ supporting an algebra $\small{\mathfrak{sp}(k)}$, where $k$ is identical to 
the vanishing order of $c$ at $t=0$.  Additionally, in most models 
there is an enhancement to a non-minimal singularity at the point $u=t=0$ which in some exceptional cases is 
shifted to $u=u_0$, $t=0$. To resolve these non-minimal points we proceed as above, i.e.\ in cycles of 
introducing base blow-ups and checking for additional non-minimal points.
%%%

To be more concrete, let us now briefly review the torically inspired blow-up procedure of \cite{Font:2016odl}. In the first step, we choose 
local affine coordinates $\xi_i$ on $B$ such that the  non-minimal singularity lies at $\xi_1=\xi_2=0$. Then, we expand the sections $f$ and $g$ in these coordinates,
\begin{equation}\label{eq:f_and_g-expansion}
f=\sum_i f_i \,\xi_1^{m^1_i}\xi_2^{m^2_i}\,,\qquad g=\sum_i g_i \,\xi_1^{l^1_i}\xi_2^{l^2_i}\,,
\end{equation}
and collect the minimal exponents $\mathbf m_i$ and $\mathbf l_i$, i.e.\ the vertices  of the Newton polytopes of $f$ and $g$ which can be connected to the origin without passing through the respective polytopes.
Next we search for all toric blow-up \cite{fulton1993introduction} directions $\mathbf n_j$ which are crepant.  
For the elliptic fibration to remain Calabi-Yau, the blow-up $\mathbf n$ must involve the fiber coordinates $x$ and $y$ too, i.e.\
\begin{equation}\label{eq:blow-up}
\xi_1,\, \xi_2,\,x,\,y\quad \mapsto 
\quad e^{n_1}\, \xi_1',\,e^{n_2}\, \xi_2',\,e^{2(n_1+n_2-1)}\, x',\,e^{3(n_1+n_2-1)}\, y' \,.
%%%
\end{equation}
Hence, the canonical class of the ambient variety after the blow-up is given by $E$ times the last column in the 
weight table
\begin{equation}
\begin{array}{|c|c|c|c|c|c|c|}
\hline
     & \xi_1 & \xi_2 & x & y & e & \sum\\\hline
 E & n_1 & n_2 & 2(n_1+n_2-1) & 3(n_1+n_2-1) & -1 &  6(n_1+n_2-1)\\ \hline
\end{array}\,,
\end{equation}
%%%
where $-E$ is the divisor class of the exceptional divisor $e=0$ and we have dropped primes to simplify notation.
Imposing that the resolution of the Weierstra\ss{} equation has to be crepant implies that
$e^{6(n_1+n_2-1)}$ must factor off the hypersurface equation \eqref{eq:Weierstrass-eqn} when its proper transform 
is taken after applying \eqref{eq:blow-up}.
This amounts then to the constraints
\begin{equation}\label{eq:allowed-blow-up-directions}
(m^1_i-4)n_1+(m^2_i-4)n_2=:\tilde{\mathbf m}_i\cdot \mathbf n\ge -4 \quad \textmd{and}\quad (l^1_i-6)n_1+(l^2_i-6)n_2=:\tilde{\mathbf l}_i\cdot \mathbf n\ge -6 
\end{equation}
which must be fulfilled for all $\tilde{\mathbf m}_i$ and $\tilde{\mathbf l}_i$. If the constraints \eqref{eq:allowed-blow-up-directions} are fulfilled for the minimal exponents then all the remaining $\mathbf m_i$'s and ${\mathbf l}_i$'s fulfill them trivially.

The set $\{{\mathbf n}^j\}$ of toric blow-ups  that need to be introduced consists of the solutions to  the 
inequalities \eqref{eq:allowed-blow-up-directions} which have coprime entries. Knowing the $\{{\mathbf n}^j\}$ it is
straightforward to compute the vanishing orders of $(f,g,\Delta)$, denoted $\mu(f)$, $\mu(g)$ and $\mu(\Delta)$,
along the corresponding exceptional divisors. This data determines the fiber type at the degeneration  \cite{kodaira123}. For the reader's convenience, we reproduced the Kodaira classification in Table~\ref{tab:Kodaira}.
The gauge algebra supported on each divisor is uniquely identified analyzing the monodromy covers \cite{Grassi:2011hq}.

\begin{table}[t]
\begin{center}
\renewcommand{\arraystretch}{1.25}
\begin{tabular}{|c|c|c|c|c|c|c|}
\hline
$\mu(f)$&$\mu(g)$&$\mu(\Delta)$ & type  &       singularity     &    gauge algebra   & monodromy \\
\hline \hline
$\geq 0$&$\geq 0$&0 &  $\kod{I_0}$   &$-$      &    $-$     
& ${\footnotesize  \begin{pmatrix} 
 1 & 0 \\ 
 0 & 1 \end{pmatrix} } $                \\ \hline
0& 0 &1 &  $\kod{I_1}$   &$-$      &    $-$      & ${\footnotesize  \begin{pmatrix} 
 1 & 1 \\ 
 0 & 1 \end{pmatrix} } $               \\ \hline
 0&0&$n$  & $\kod{I}_n$      &     $\mathrm{A}_{n-1}$     &  $\mathfrak{su}(n)$  or $\mathfrak{sp}([n/2])$ 
 & ${\footnotesize  \begin{pmatrix} 
 1 & n \\ 
 0 & 1 \end{pmatrix} } $     \\ \hline
   $\geq 1$ &1&2 & $\kod{II}$ &\text{cusp}       &    $-$  
& ${\footnotesize  \begin{pmatrix} 
 1 & 1 \\ 
 -1 & 0 \end{pmatrix} } $     \\ \hline
  1 &$\geq 2$ & 3     & $\kod{III}$ &    $\mathrm{A}_{1}$  &    $\mathfrak{su}(2)$   
& ${\footnotesize  \begin{pmatrix} 
 0 & 1 \\ 
 -1 & 0 \end{pmatrix} } $    \\ \hline
$\geq 2$ &2 & 4   & $\kod{IV}$    &    $\mathrm{A}_{2}$    &    $\mathfrak{su}(3)$  or $\mathfrak{sp}(1)$  
& ${\footnotesize  \begin{pmatrix} 
 0 & 1 \\ 
 -1 & -1 \end{pmatrix} } $  \\ \hline
$\geq2 $&$\geq 3$&6     &  $\kod{I_0^{\ast}}$&  $\mathrm{D}_4$  &  $\mathfrak{so}(8)$  or $\mathfrak{so}(7)$ or $\mathfrak{g}_2$ 
& ${\footnotesize  \begin{pmatrix} 
 -1 & 0 \\ 
 0 & -1 \end{pmatrix} } $\\ \hline
2&3&$n+6$    & $\kod{I}_n^{\ast}$ &  $\mathrm{D}_{4+n}$   &  $\mathfrak{so}(2n\text{+}8)$  or $\mathfrak{so}(2n\text{+}7)$   
& ${\footnotesize  \begin{pmatrix} 
 -1 & -n \\ 
 0 & -1 \end{pmatrix} } $  \\ \hline
$\geq 3$ &4&8    & $\kod{IV^{\ast}}$&      $\mathrm{E}_6$    &    $\mathfrak{e}_6$   or $\mathfrak{f}_4$     
& ${\footnotesize  \begin{pmatrix} 
 -1 & -1 \\ 
 1 & 0 \end{pmatrix} } $\\ \hline
   3  &$\geq 5$ & 9     & $\kod{III^{\ast}}$&   $\mathrm{E}_7$  &    $\mathfrak{e}_7$  
   & ${\footnotesize  \begin{pmatrix} 
 0 & -1 \\ 
 1 & 0 \end{pmatrix} } $\\ \hline
$\geq 4$ &5&10     & $\kod{II^{\ast}}$ &     $\mathrm{E}_8$    &   $\mathfrak{e}_8$     
 & ${\footnotesize  \begin{pmatrix} 
 0 & -1 \\ 
 1 & 1 \end{pmatrix} } $        \\ \hline 
\end{tabular}
\caption{Kodaira classification of degenerate elliptic fibers.}
  \label{tab:Kodaira}\end{center}\end{table}

After this toric resolution step we still have to check whether there are no non-minimal points along the exceptional 
curves just introduced. If there are any of them, we must repeat the resolution procedure, which we just described, at these points. 
The process of resolving and checking stops when all non-Kodaira type singularities have been removed.

As realized in \cite{Font:2016odl}, there might be cases when the resolution cannot be accomplished. This occurs
when there is an infinite number of solutions to \eqref{eq:allowed-blow-up-directions}. Moreover, it can be shown that the
set of blow-ups is finite  if and only if  $\mu(a)<4$ or $\mu(b)<6$ or $\mu(c)<10$ or $\mu(d)<12$. 

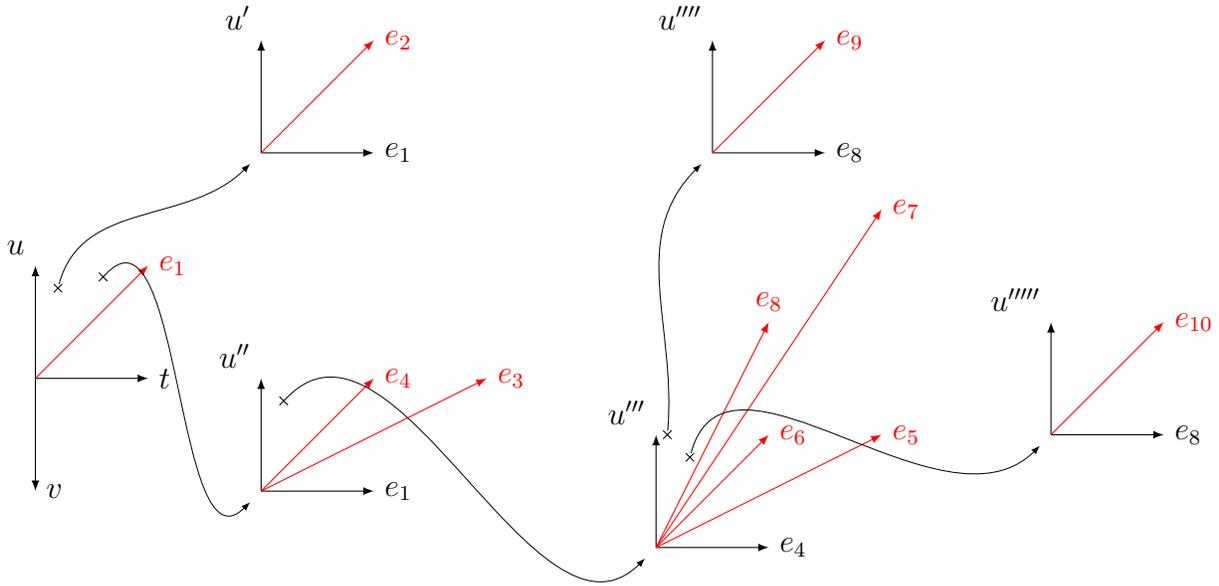
\begin{figure}
\begin{tikzpicture}[>=latex,
res vec/.style={red,->},
cross/.style={cross out, draw=black, fill=none, minimum size=2*(#1-\pgflinewidth), inner sep=0pt, outer sep=0pt}, cross/.default={2pt},
pfeile/.style={very thin,->}]
\begin{scope}[name=res0,scale=1.5]
%%%
\node at (0,0) (zero) {};
 \draw[->] (zero.center) -- ++(1,0) node[right] {$t$};
 \draw[->] (zero.center) -- ++(0,1) node[above left] {$u$};
 \draw[->] (zero.center) -- ++(0,-1) node[right] {$v$};
%%%
 \draw[res vec] (zero.center) -- (1,1) node[right] {$e_1$};
%%%
\node[cross] (nmp1) at (0.2,0.8) {}; 
\node[cross] (nmp2) at (0.6,0.9) {}; 
\end{scope}
%%%
\begin{scope}[name=res1,scale=1.5]
 \path (2,2) node (zero) {};
 %%%
 \draw[->] (zero.center) -- ++(1,0) node[right] {$e_1$};
 \draw[->] (zero.center) -- ++(0,1) node[above left] {$u'$};
%%%
 \draw[res vec] (zero.center) -- ++(1,1) node[right] {$e_2$};
\end{scope}
\draw[pfeile] (nmp1) to[out=75,in=225] (zero);
%%%
\begin{scope}[name=res2,scale=1.5]
 \path (2,-1) node (zero) {};
 %%%
 \draw[->] (zero.center) -- ++(1,0) node[right] {$e_1$};
 \draw[->] (zero.center) -- ++(0,1) node[above left] {$u''$};
%%%
 \draw[res vec] (zero.center) -- ++(2,1) node[right] {$e_3$};
 \draw[res vec] (zero.center) -- ++(1,1) node[right] {$e_4$};
%%%
\path (zero.center) -- ++ (0.2,0.8) node[cross] (nmp3) {}; 
\end{scope}
\draw[pfeile] (nmp2) to[out=45,in=225] (zero);
%%%
\begin{scope}[name=res3,scale=1.5]
 \path (5.5,-1.5) node (zero) {};
 %%%
 \draw[->] (zero.center) -- ++(1,0) node[right] {$e_4$};
 \draw[->] (zero.center) -- ++(0,1) node[above left] {$u'''$};
%%%
 \draw[res vec] (zero.center) -- ++(2,1) node[right] {$e_5$};
 \draw[res vec] (zero.center) -- ++(1,1) node[right] {$e_6$};
 \draw[res vec] (zero.center) -- ++(2,3) node[right] {$e_7$};
 \draw[res vec] (zero.center) -- ++(1,2) node[above] {$e_8$};
%%%
\path (zero.center) -- ++ (0.1,1) node[cross] (nmp4)  {}; 
\path (zero.center) -- ++ (0.3,0.8) node[cross] (nmp5) {}; 
\end{scope}
\draw[pfeile] (nmp3) to[out=45,in=225] (zero);
%%%
\begin{scope}[name=res4,scale=1.5]
 \path (6,2) node (zero) {};
 %%%
 \draw[->] (zero.center) -- ++(1,0) node[right] {$e_8$};
 \draw[->] (zero.center) -- ++(0,1) node[above left] {$u''''$};
%%%
 \draw[res vec] (zero.center) -- ++(1,1) node[right] {$e_9$};
\end{scope}
\draw[pfeile] (nmp4) to[out=85,in=225] (zero);
%%%
\begin{scope}[name=res5,scale=1.5]
 \path (9,-.5) node (zero) {};
 %%%
 \draw[->] (zero.center) -- ++(1,0) node[right] {$e_8$};
 \draw[->] (zero.center) -- ++(0,1) node[above left] {$u'''''$};
%%%
 \draw[res vec] (zero.center) -- ++(1,1) node[right] {$e_{10}$};
\end{scope}
\draw[pfeile] (nmp5) to[out=75,in=225] (zero);
\end{tikzpicture}
 \caption{Schematic drawing of the toric resolutions of the $\mathrm{II}_{9-6}$ singularity, cf.~Section~\ref{sec:parabolic-type-4}. The resolution starts on the left hand side with the base blow-up of $u=t=0$. After this blow-up there are two non-minimal points in the $u$-$e_1$-patch which are indicated by the two crosses. Both singularities lie along $e_1=0$ but $u$ non-vanishing. Therefore, we have to do a coordinate redefinition such that the singularities lie at $e_1=u'=0$ and $e_1=u''=0$, respectively, to apply again our toric machinery. These cycles of toric resolution and searching for non-minimal points continue in the obvious way until all non-minimal singularities are removed.}\label{fig:resolutions}
\end{figure}

To obtain the self-intersection numbers $a_i$ of the (blow-up) divisors in the base, it is too na\"ive to take the respective toric resolution and calculate $a_i$ from
\begin{equation}
 \mathbf n^{i+1}+\mathbf n^{i-1}=a_i\,\mathbf n^i
\end{equation}
where the $\mathbf n^i$'s are the lattice vectors corresponding to blow-up divisors. 
%%%
The reason for this is that when we do the cycles of toric resolutions and checking for non-minimal points we change the self-intersection number of the (toric) divisors on which we still find non-minimal points after the toric resolution, because we have to blow-up these points in the next cycle. The self-intersection of the divisor changes by $-1$ for each non-minimal point which lies on that divisor and we have to blow-up. Note that this is not only true for the blow-up divisors but also for the rational curve at $t=0$. Although it has self-intersection number $0$ in the beginning, its self-intersection number becomes $-1$ after resolving the non-minimal point at $t=u=0$,\footnote{Since the divisor $u=0$ is non-compact, we cannot define a self-intersection number and, therefore, no change in it.} cf.~Figure~\ref{fig:resolutions}.

\subsection{Geometric Models: Small Instantons on ADE Singularities}
\label{sec:ADE}

In this section we discuss resolutions of $\mathrm{Spin}(32)/\ZZ_2$ models 
which on the genus-two side have a NU degeneration $[\mathrm{I}_{n-p-0}]$,
$[\mathrm{I}_n-\mathrm{I}^*_p]$ and $[\mathrm{K}-\mathrm{I}_n]$, with $\mathrm{K}=\mathrm{II}^*, \mathrm{III}^*, \mathrm{IV}^*$  \cite{Namikawa:1973yq}.
Here we use the notation $\kod{[K_1-K_2-0]}\equiv \kod{[K_1-K_2]}$.
These models are expected to correspond to heterotic compactifications with small instantons sitting at
ADE singularities  based on the monodromy action on the moduli and the Bianchi identity 
$dH \sim (\mathrm{Tr}\, F^2 - \mathrm{Tr}\, R^2)$.
For example, in the $[\mathrm{II}^{\ast}-\mathrm{I}_n]$ model the monodromy is
\begin{equation}
\label{monoe8}
\tau \rightarrow -\frac{1}{1+\tau} \, ,\quad \rho \rightarrow \rho + n - \frac{\beta^2}{1+\tau}
\, ,\quad \beta \rightarrow
\frac{\beta}{1+\tau}\, .
\end{equation}
When the Wilson line value $\beta$ is turned off, $\rho \to \rho + n$, whereas the monodromy in $\tau$
is precisely that of a Kodaira type $\kod{II^{\ast}}$ fiber of the $\tau$ fibration. Indeed, shortly we will see that
the model describes $k=10+n$ instantons on an ${\mathrm E}_8$ singularity.

As explained in Section~\ref{ss:ftheory}, the starting point is the
genus-two model given in the NU classification. The next step is to compute the Igusa-Clebsch invariants that determine 
the $a,b,c$ and $d$ coefficients entering in the dual K3 on the F-theory side. In Table~\ref{tab:adetype} we collect the 
defining equations of the ADE NU models together with the vanishing degrees at $t=0$ of the coefficients $a,b,c$ and $d$. From the latter we can infer the behavior of the functions $Q$ and $\epsilon$, cf.~\eqref{qedef}, which control the loci of 
singularities and small instantons on the heterotic side \cite{Aspinwall:1996nk, Aspinwall:1996vc, Aspinwall:1997ye}. 
In particular, it follows that there are $k$ small instantons on top of the $\mathrm{K}$-type singularity, where 
\begin{equation}
\label{kinst}
k=\mu(c)
 \end{equation}
is precisely the vanishing degree of $\epsilon$ at $t=0$.

\begin{table}[h!]\begin{center}
\renewcommand{\arraystretch}{1.5}
{\small
\begin{tabular}{|M{.8cm}|M{1.5cm}|M{6.5cm}|c|c|c|c|}
\hline
sing. & NU   type       &  local model &  $\mu(a)$ & $\mu(b)$ &$\mu(c)$& $\mu(d)$        \\[3pt]
\hline
$\mathrm{A}_{p-1}$ &  $[\mathrm{I}_{n-p-0}]$  &$ y^2\!=\!\left(t^n+x^2\right) \left(t^p+(x-\alpha)^2\right)\left(x-1\right)  $& $0$&  $0$&$n+p$ & $n+p$  
  \\[3pt]\hline
$\mathrm{D}_{p+4}$ & $[\mathrm{I}_n-\mathrm{I}^*_p]$  & $y^2\!=\!\left(t^n+(x-1)^2\right) \left(t^{p+2}+x^2\right)\left(x+t\right)  $& $2$&  $3$&$6+n+p$ & $6+n+p$  
  \\[3pt]\hline
$\mathrm{E}_6$ & $[\mathrm{IV}^{\ast}-\mathrm{I}_n]$  &$y^2\!=\!\left(t^{4}+x^3\right) \left(t^n+(x-1)^2\right) $& $4+n$&  $4$&$8+n$ & $8+n$  
  \\[3pt]\hline
$\mathrm{E}_7$ & $[\mathrm{III}^{\ast}-\mathrm{I}_n]$  &$y^2\!=\!x\left(t^{3}+x^2\right) \left(t^n+(x-1)^2\right) $& $3$&  $6+n$&$9+n$ & $9+n$  
  \\[3pt]\hline
$\mathrm{E}_8$ &   $[\mathrm{II}^{\ast}-\mathrm{I}_n]$  &$y^2\!=\!\left(t^{5}+x^3\right) \left(t^n+(x-1)^2\right) $& $5+n$&  $5$&$10+n$ & $10+n$  
  \\[3pt]\hline
\end{tabular}
}
\caption{Genus-two models for ADE singularities.}
  \label{tab:adetype}\end{center}\end{table}

On the F-theory side there are non-minimal points which we seek to resolve following the procedure described in Section~\ref{ss:toric}.
In general the resolution consists of a series of base blow-ups. Each divisor can be
characterized by an integer equal to minus its self-intersection number, and by the
gauge algebra factor it supports. This algebra is derived from the vanishing orders of $(f,g,\Delta)$ along the blow-up divisors,
cf.~Table~\ref{tab:Kodaira}, and the study of the monodromy covers following the formalism of \cite{Grassi:2011hq}. 
In order to determine the matter content it is also important
to give the intersection pattern of the blow-ups. All this information can be efficiently found 
using the toric geometry techniques reviewed in the preceding section.

In the end, to each model admitting a resolution we can associate a full local gauge
algebra, denoted $\mathcal G$, and a number of blow-ups, denoted $n_T$. In turn $n_T$ counts the
massless tensor multiplets \cite{Aspinwall:1997ye}.  In all models the full algebra and the 
total number of blow-ups agree with the results obtained originally in  \cite{Aspinwall:1997ye}. We complete
these results by providing the  complete pattern of the curves supporting the algebras.
In fact, each pattern fits the predictions based on the analysis of the theory of small $SO(32)$ instantons on
$\CC^2/\Gamma_G$ singularities, where $\Gamma_G$ is the discrete subgroup of $SU(2)$ 
associated to the ADE group $G$ \cite{Blum:1997mm}. More precisely, for a singularity of type $G$ the structure of the 
resolution is dictated by the extended Dynkin diagram of $G$. 
At the nodes of the diagram,  labelled by $\nu=0, 1,\cdots, {\text {rank}}\,  G$,  
there are algebras of type $\mathfrak{sp}(v_\nu)$, $\mathfrak{so}(v_\nu)$, or $\mathfrak{su}(v_\nu)$, 
according to whether the representation of $\Gamma_G$ associated to the respective node is real, pseudoreal, or complex. 
A node with the conjugate of a complex  representation does not give a new algebra factor. 
For example, for an  $\mathrm{A}_{p-1}$ singularity with $p$ odd, the resolution has one 
$\mathfrak{sp}$ and $(p-1)/2$ $\mathfrak{su}$ factors because there is only one
real, together with $(p-1)/2$ complex nodes plus their conjugates.  The values of the $v_\nu$ depend on 
data of the extended diagram. We refer to \cite{Blum:1997mm} for details.
In particular, it follows that the extended node of the diagram, which is always real, gives an algebra $\mathfrak{sp}(k)$, where
$k$ is the number of instantons on the singularity. 
This is just the factor due to the singularity at $t=0$ in the ADE NU degenerations.
%%%
%%%
%%%
%%%
%%%
%%%

The matter content can also be determined from the resolution output and it agrees with the predictions in \cite{Blum:1997mm} as well.  Concretely, each $\mathfrak{sp}(v_\nu)$, $\mathfrak{so}(v_\nu)$, $\mathfrak{su}(v_\nu)$, sitting 
at a curve of self-intersection $-1$, $-4$, $-2$, respectively, has altogether $2v_\nu + 8$, $v_\nu - 8$, $2v_\nu$ fundamentals.
A $\mathfrak{su}(v_\nu)$ at a curve of self-intersection $-1$ has $v_\nu + 8$ fundamentals plus
an extra hypermultiplet in the antisymmetric representation.
Also in other cases, it can be shown that the matter necessary for anomaly cancellation is present.  For instance, if we take the $\mathfrak{so}(7)$
at a curve of self-intersection $-2$, the analysis of the monodromy cover indicates that there is one hypermultiplet in the fundamental
and four in the spinor representation.

\begin{figure}
\begin{minipage}{.3\textwidth}
\begin{flushleft}
\begin{tikzpicture}[scale=.2]
%%%
 \draw (15*\PicScale,18*\PicScale) node[left]  {a) $\mu(d)=\mu(c)$};
%%%
\draw[thick] (-5*\PicScale, 0) -- (18*\PicScale, 0);
 \draw (18.5*\PicScale, 0) node[right]  {};
%%%
\draw[thick] (-4*\PicScale, -1*\PicScale) -- (-4*\PicScale, 13*\PicScale);
  \draw (-4*\PicScale, -1.1*\PicScale) node[below]  {\msm{t=0}};
    \draw (-4*\PicScale, 12.5*\PicScale) node[above right]  {\msm{\mathrm{I}_{2k}}};
%%%
   \draw[thick] (-8*\PicScale, 10*\PicScale) -- (4*\PicScale, 10*\PicScale);
      \draw (-8*\PicScale, 10*\PicScale) node[left]  {\msm{v=0}};
            \draw (3.25*\PicScale, 10*\PicScale) node[above right]  {\msm{\mathrm{I}^*_{10}}};
%%%
  \draw[thick] (-6.5*\PicScale, 6*\PicScale) -- (2.5*\PicScale, 6*\PicScale);
      \draw (-6.5*\PicScale, 6*\PicScale) node[left]  {\msm{v=-\dfrac{\tilde c}{\tilde d}}};
            \draw (1.85*\PicScale, 6*\PicScale) node[above right]  {\msm{\mathrm{I}_2}};
%%%
  \draw[thin]  (8*\PicScale, -0.8*\PicScale) -- (8*\PicScale, 3*\PicScale);
    \draw[thin]  (7.5*\PicScale, 2*\PicScale) -- (12*\PicScale, 2*\PicScale);
        \draw[thin]  (11.5*\PicScale, 1.5*\PicScale) -- (11.5*\PicScale, 5.5*\PicScale);
           \draw[thin]  (11*\PicScale, 4.5*\PicScale) -- (15*\PicScale, 4.5*\PicScale);
       \draw[thin]  (14.5*\PicScale, 4*\PicScale) -- (14.5*\PicScale, 8*\PicScale);
 \draw (14.5*\PicScale, 7.5) node[right]  {\msm{\dots}};
    \end{tikzpicture}
    \end{flushleft}
\end{minipage}%%%
\hspace{4cm}
\begin{minipage}{.3\textwidth}
\begin{flushright}
\begin{tikzpicture}[scale=.2]
%%%
 \draw (15*\PicScale,18*\PicScale) node[left]  {b) $\mu(d) >\mu(c)$};
%%%
\draw[thick] (-5*\PicScale, 0) -- (18*\PicScale, 0);
 \draw (18.5*\PicScale, 0) node[right]  {};
%%%
\draw[thick] (-4*\PicScale, -1*\PicScale) -- (-4*\PicScale, 13*\PicScale);
  \draw (-4*\PicScale, -1.1*\PicScale) node[below]  {\msm{t=0}};
    \draw (-4*\PicScale, 12.5*\PicScale) node[above right]  {\msm{\mathrm{I}_{2k}}};
%%%
   \draw[thick] (-8*\PicScale, 10*\PicScale) -- (4*\PicScale, 10*\PicScale);
      \draw (-8*\PicScale, 10*\PicScale) node[left]  {\msm{v=0}};
            \draw (3.25*\PicScale, 10*\PicScale) node[above right]  {\msm{\mathrm{I}^*_{10}}};
%%%
  \draw[thick] (10*\PicScale, 3*\PicScale) -- (14*\PicScale, 3*\PicScale);
            \draw (13.85*\PicScale, 1.5*\PicScale) node[above right]  {\msm{\mathrm{I}_2}};
%%%
  \draw[thin]  (8*\PicScale, -0.8*\PicScale) -- (8*\PicScale, 3*\PicScale);
    \draw[thin]  (7.5*\PicScale, 2*\PicScale) -- (12*\PicScale, 2*\PicScale);
        \draw[thin]  (11.5*\PicScale, 1.5*\PicScale) -- (11.5*\PicScale, 5.5*\PicScale);
           \draw[thin]  (11*\PicScale, 4.5*\PicScale) -- (15*\PicScale, 4.5*\PicScale);
       \draw[thin]  (14.5*\PicScale, 4*\PicScale) -- (14.5*\PicScale, 8*\PicScale);
 \draw (14.5*\PicScale, 7.5) node[right]  {\msm{\dots}};
    \end{tikzpicture}
    \end{flushright}
\end{minipage}%%%
\caption{Schematic representation of blow-ups and intersections of characteristic divisors}\label{bild}
%%%
\end{figure}
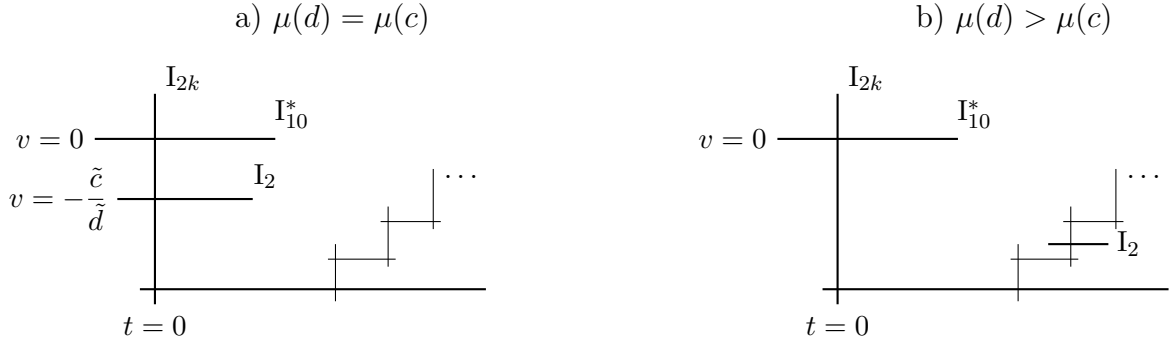

There is also matter due to intersections of the global $\mathrm{Spin}(28) \times SU(2)/\ZZ_2$ with the local base blow-ups and the rational curve at $t=0$
as we now explain. The discriminant \eqref{discri32}  shows that the 
fiber $\mathrm{I}^*_{10}$ ($\mathfrak{so}(28)$) is at $v=0$ whereas the $\mathrm{I}_2$ ($\mathfrak{su}(2)$) occurs at 
the vanishing of $(cu+dv)$. Recall further that from the term $(cu+dv)^2$ there factors off the $\mathrm{I}_{2k}$ ($\mathfrak{sp}(k)$) at $t = 0$, with $k = \mu(c)$.
At lowest order in $t$ we can write $c=t^{\mu(c)} \tilde c$ and $d=t^{\mu(d)} \tilde d$, where $\tilde c$ and $\tilde d$ are some
non-zero constants. In the case that $\mu(d)=\mu(c)$, as in the examples in Table~\ref{tab:adetype}, 
the locus of the $\mathrm{I}_2$ fiber intersects the $\mathrm{I}_{2k}$ singularity at $t=0,\,v=-\tilde c/\tilde d$, in the patch $u=1$.
This is in contrast to the case $\mu(d) > \mu(c)$, then the intersection locus will be at $u = t = 0$ where we have to introduce the base blow-ups. %%%
 As it turns out, the intersection pattern of the $\mathrm{I}_2$-locus with the blow-up divisor is model-dependent. The upshot is that in either case the $\mathfrak{so}(28)$ only intersects the $\mathfrak{sp}(k)$ and there
is matter $\frac12({\bf{28}}, {\pmb{2 k}})$ at this point. Concerning the $\mathfrak{su}(2)$ locus, for $\mu(d) = \mu(c)$ 
it intersects $\mathfrak{sp}(k)$ and there is an additional  $({\bf{2}}, {\pmb{2 k}})$, whereas 
for $\mu(d) > \mu(c)$ it will intersect one of the divisors introduced by the resolution of the non-minimal singularity at $u=t=0$.  The two situations are depicted in Figure~\ref{bild}.
In all examples, including matter from all intersections, the $\mathfrak{sp}(k)$ has altogether $2k+8$ hypermultiplets in the
fundamental representation.

We will display the results using  a notation such that each blow-up divisor introduced in the resolution
is identified by the algebra it supports written above an integer which is equal to minus its self-intersection 
number. In the  $\mathrm{Spin}(32) /\ZZ_2$ heterotic, below the universal $\mathfrak{sp}(k)$, supported along the curve $t=0$ with self-intersection $-1$,
we will write $1^*$, adding the asterisk to indicate that $t=0$ is not a blow-up divisor. 
Besides, adjacent divisors intersect and when necessary this is made clear by drawing an explicit link.
Thus, a generic point on the tensor branch of the 6d ${\mathcal N}=(1,0)$ theory corresponding to a resolvable degeneration
will be captured by a tree-like diagram with  $n_T+1$ nodes. 

Anomaly cancellation gives significant information about the resulting 6d ${\mathcal N}=(1,0)$ theories. In all models it happens that
the matter content is such that the irreducible gauge quartic anomaly cancels for each gauge factor. Moreover, the remaining pure
gauge contribution to the anomaly polynomial takes the form{\footnote{We use the conventions of \cite{Ohmori:2014kda}
for the anomaly polynomial and those of \cite{Grassi:2011hq} for the traces involved.}}
\begin{equation}
\label{gsfact}
I_{\text{gauge}}= -\frac18 \eta^{\alpha\beta} \, \text{tr} F_\alpha^2  \, \text{tr} F_\beta^2\,  .
\end{equation} 
Here $F_\alpha$ is the field strength of the gauge factor at 
the $\alpha^{\mathrm{th}}$ node, with $\alpha=0,\, 1,\ldots,\,n_T$, where $\alpha=0$ refers to $\mathfrak{sp}(k)$,
and the so-called adjacency matrix $\eta^{\alpha\beta}$ is equal to minus the self-intersection
matrix. If there is no algebra at the node we set $F_\alpha=0$.
The adjacency matrix can be read off from the diagrams representing the theories, for an example see e.g.~\eqref{newE8}. 
Concretely, the diagonal elements of $\eta^{\alpha\beta}$ are the integers under the nodes in the diagram while the off-diagonal
elements are $-1$ or 0 depending on whether the nodes are linked or not.
In all models one can check that $\eta^{\alpha\beta}$ is positive 
semi-definite, with only one zero eigenvalue. In consequence, $I_{\text{gauge}}$ can be cancelled by the Green-Schwarz-Sagnotti
mechanism \cite{Green:1984sg, Sagnotti:1992qw} involving just $n_T$ tensor multiplets \cite{Blum:1997mm}. 
The null eigenvalue further implies that a linear combination of gauge couplings is independent of the scalars in the tensor 
multiplets and therefore it defines a mass parameter.
%%%
%%%
%%%

The existence of a mass scale suggests that the UV completion of the theories arising from the resolutions are little string theories (LSTs)
\cite{Seiberg:1997zk}. In fact, the theories that we obtain have appeared in the recent classifications of LSTs \cite{Bhardwaj:2015xxa, Bhardwaj:2015oru}. Moreover, 
dropping the node corresponding to $t=0$ in the diagrams, i.e.\ deleting the corresponding column and row in $\eta^{\alpha\beta}$,
gives the tensor branch of 6d SCFTs embedded in the LSTs \cite{Bhardwaj:2015oru}. In this case the $\mathfrak{sp}(k)$
remains as a flavor symmetry of the 6d SCFTs as observed originally in \cite{Blum:1997mm}. In all cases we find that 
the residual $n_T \times n_T$ adjacency matrix, denoted $\eta^{ij}$, $i,j=1,\ldots,\,n_T$, is positive definite, has determinant one,
and further satisfies
\begin{equation}
\label{etaprop}
\sum_{i=1}^{n_T} \sum_{j=1}^{n_T} \left(\eta^{-1}\right)_{ij} \left(2 - \eta^{ii}\right) \left(2 - \eta^{jj}\right) = n_T \, .
\end{equation}
This property enters in the computation of the anomaly polynomial of the SCFTs applying the methods developed in
\cite{Ohmori:2014kda, Intriligator:2014eaa}.

An interesting feature of the theories emerging from the resolution of NU degenerations is that they can be characterized
by some quantities that match in the $\mathrm{Spin}(32)/\ZZ_2$ and the $E_8 \times E_8$ heterotic strings. 
For instance, for a concrete degeneration with resolution $\mathrm R$, the quantity 
\begin{equation}
\label{checkdef}
h_{\mathrm R}={\text {rank}}\,  {\mathcal G} + n_T \, , 
\end{equation}
can be shown to be the same for both heterotic strings by virtue of duality upon further compactification on a circle
\cite{Aspinwall:1997ye}. We have found that this indeed occurs, which actually provides an useful check of the results.
Moreover, for the particular case of the models
in Table~\ref{tab:adetype}, corresponding to $k=\mu(c)$ small instantons on ADE singularities, it turns out that for
$k$ above a minimum value the resolution satisfies 
\begin{equation}
\label{checkade}
h_{\mathrm R}= g_G k - {\text {dim}}\, G , 
\end{equation}
where $g_G$ is the Coxeter number of the ADE group $G$, given by $g_G=p,\,2p+6,\,12,\,18,\,30$, for
$G=SU(p),\, SO(2p+8),\, E_6,\, E_7,\, E_8$, respectively. This fact was observed in  \cite{Perevalov:1997ht}.

To each resolution we can assign a second intrinsic quantity that takes the same value for both heterotic strings.
Knowing the local algebra ${\mathcal G}$ and the matter content it is easy to compute the number of
vector multiplets given by $n_V={\text {dim}}\,  {\mathcal G}$ and the total number of hypermultiplets $n_H$.
The number of tensor multiplets $n_T$ and the instanton number $k=\mu(c)$ are also inherent properties of the theory 
derived from the concrete resolution. With this data we define
\begin{equation}
\label{checkrank}
r_{\mathrm R}=n_H - n_V +29 n_T -30 k\, . 
\end{equation}
An indication that $r_{\mathrm R}$ depends only on the underlying NU degeneration, so it matches in both heterotic strings,
is the fact that in all models corresponding to small instantons on ADE singularities $r_{\mathrm R}=\mathrm {rank}\, G$,
as pointed out in \cite{Intriligator:1997dh}. 
One way to derive the relation \eqref{checkrank} 
is to consider a global heterotic model constructed as a compactification on K3 with $(24-k)$ large $SU(2)$ instantons
breaking the gauge group to $\mathrm{Spin}(28) \times SU(2)/\ZZ_2$ or $E_7 \times E_8$ \cite{Green:1984bx}, plus $k$ small
instantons on the ADE singularity giving the local ${\mathcal G}$ theory.
Imposing cancellation of the pure gravitational anomaly leads to \eqref{checkrank}.  
%%%
%%%

It is worthwhile to compare the resolutions of the same NU model in both heterotic strings, for instance to
check the matching of the quantities $h_{\mathrm R}$ and $r_{\mathrm R}$ defined above.
To this end, we will give in the current section the resolutions in the $E_8\times E_8$ string too. In the diagrams representing
the resulting theories we will also include the $t=0$ divisor with label $1^*$, 
but in the $E_8\times E_8$ string it does not support any gauge algebra.
The pure gauge anomaly in the resulting theories again takes the form \eqref{gsfact}. In all cases the self-intersection matrix
$\eta^{\alpha\beta}$ is positive semi-definite with a single null eigenvalue.
Hence, also these theories potentially complete to LSTs in the UV. 
Similar claims have been made in \cite{Bhardwaj:2015oru} for the theories associated to small instantons on 
ADE singularities that we consider in this section. Notice that T-duality upon circle compactification, reflected in the double
fibration structure of the F-theory duals, requires that the resulting theories in both heterotic strings be LSTs \cite{Bhardwaj:2015oru}. %%%
Again, dropping the $t=0$ node gives the tensor branch description of 6d SCFTs embedded in the LSTs. 
%%%
%%%
This is the situation which was implicitly assumed in \cite{Font:2016odl}.

Below we will present the resolutions of three examples of Table~\ref{tab:adetype} which are relevant for the
ensuing discussion. The remaining  models can be found in  Appendix~\ref{app:ADE}. 

\subsubsection{\texorpdfstring{$\kod{[II^{\ast}-I_n]}$}{[II^*-I_n]} Model and \texorpdfstring{$\mathrm{E}_8$}{E8} Singularity}\label{ss:InE8}

The number of small instantons on the $E_8$ singularity is $k=10+n$. For $n \ge 1$ the resolution 
in the $E_8 \times E_8$ heterotic string gives
\begin{align}
\label{resoInE8}
&
\footnotesize{
\begin{tabular}{|cccccccccc|}
\hline
& & $\mathfrak{sp}(1)$ &$\mathfrak{g}_2$& & $\mathfrak{f}_4$& &  $\mathfrak{g}_2$& $\mathfrak{sp}(1)$ &  \\
1 & 2& 2& 3& 1& 5 & 1 & 3 & 2 & 2 \\
\hline
\end{tabular}\, 
}
\footnotesize{
\begin{tabular}{|cccccccccccc|}
\hline
 & 1 &  & & & & & &  &  &  & \\
 & $\mid$ &  & & & & & &  &  &  &  \\
 & $\mathfrak{e}_8$& & & $\mathfrak{sp}(1)$ &$\mathfrak{g}_2$& & $\mathfrak{f}_4$& &  $\mathfrak{g}_2$& $\mathfrak{sp}(1)$ &  \\
1 & 12 & 1 & 2& 2& 3& 1& 5 & 1 & 3 & 2 & 2 \\
\hline
\end{tabular}\, 
\times  }   \nonumber  \\ 
& 
\footnotesize{
\begin{tabular}{|cccccccccccc|}
\hline
 & $\mathfrak{e}_8$& & & $\mathfrak{sp}(1)$ &$\mathfrak{g}_2$& & $\mathfrak{f}_4$& &  $\mathfrak{g}_2$& $\mathfrak{sp}(1)$ &  \\
1 & 12 & 1 & 2& 2& 3& 1& 5 & 1 & 3 & 2 & 2 \\
\hline
\end{tabular}^{\, \oplus (n-1)}
\hspace*{-5mm}  \times} \\ \nonumber
& 
\footnotesize{ \times
\begin{tabular}{|ccccccccccccc|}
\hline
 & $\mathfrak{e}_8$& & & $\mathfrak{sp}(1)$ &$\mathfrak{g}_2$& & $\mathfrak{f}_4$& &  $\mathfrak{g}_2$& $\mathfrak{sp}(1)$ &  &\\
1 & 12 & 1 & 2& 2& 3& 1& 5 & 1 & 3 & 2 & 2 & $1^*$\\
 & $\mid$ &  &  &  &  &  &   &   &   &   &   &\\
  & 1         &  &  &  &  &  &   &   &   &   &   &\\
\hline
\end{tabular}\, 
\, . }
\end{align}
A systematic analysis reveals that at both the leftmost and rightmost divisors 
with singular type ${\mathrm {II}}^*$ fiber, supporting algebra $\mathfrak{e}_8$, there is one additional 
non-minimal point which requires an extra blow-up with ${\mathrm I}_0$ fiber and hence no algebra. 
In  \cite{Font:2016odl} these $\mathfrak{e}_8$ divisors were reported with self-intersection $-11$.
%%%
However, it is understood that a single curve with $\mathfrak{e}_8$ algebra and self-intersection $-11$ comes with one small instanton \cite{Heckman:2015bfa}. %%%
The resolution shown in \eqref{resoInE8}  makes this explicit. %%%
%%%
Similarly, one can readily verify that $r_{\mathrm R}=8$, since the only matter are hypermultiplets 
transforming as $\frac12(\mathbf{2},\mathbf{1}) \oplus \frac12(\mathbf{2},\mathbf{7})$ in each
$\mathfrak{sp}(1) \oplus \mathfrak{g}_2$ cluster.

In the $\mathrm{Spin}(32)/\ZZ_2$ heterotic string, for $n \ge 1$, $k \ge 11$, we obtain the resolutions
\begin{equation}
\footnotesize{
\label{newE8}
\begin{tabular}{|cccccccc|}
\hline
 & & & & & $\mathfrak{sp}(3k\text{-}32)$ & & \\
 & & & & & 1 & & \\
 & & & & & $\mid$ & & \\
$\mathfrak{sp}(k)$ &  $\mathfrak{so}(4k\text{-}16)$ &$\mathfrak{sp}(3k\text{-}24)$& $\mathfrak{so}(8k\text{-}64)$&
  $\mathfrak{sp}(5k\text{-}48)$&$\mathfrak{so}(12k\text{-}112)$&$\mathfrak{sp}(4k\text{-}40)$&
 $\mathfrak{so}(4k\text{-}32) $ \\ 
1*&4&1&4&1&4&1&4 \\\hline
\end{tabular} \, .
}
\end{equation}
Notice that the structure of the intersections 
mimics the extended Dynkin diagram of $E_8$, in agreement with the analysis of \cite{Blum:1997mm}. 
The algebra factor $\mathfrak{sp}(k)$ arises from the singularity at $t=0$.  
The total number of base blow-ups is $n_T=8$ and the rank of the full
algebra is such that $h_{\mathrm R}= 30 k - 248$, which is also the value obtained for the resolution in \eqref{resoInE8}.
It is also straightforward to check that $r_{\mathrm R}= 8$ because the matter hypermultiplets 
comprise $\frac12(\mathbf{fund}, \mathbf{fund})$ for each pair of adjacent $\mathfrak{sp}$-$\mathfrak{so}$
algebras.
%%%
%%%
Besides, there are 16 additional  fundamentals of $\mathfrak{sp}(k)$ arising from the intersection with the global 
$\mathrm{Spin}(28) \times SU(2)/\ZZ_2$, as explained above.

Let us now consider the case $n=0$, $k=10$, which will be important for latter purposes.
In the $E_8 \times E_8$ string the resolution yields
\begin{equation}
\label{newE8n0}
\footnotesize{
\begin{tabular}{|ccccccccccccccccccccccc|}
\hline
  & &  & & & & & & & & & 1 & & & & & & & & & &  &\\
 & &  & & & & & & & & &$\mid$ & & & & & & & & & & & \\
 &  &$\mathfrak{sp}(1)$ &$\mathfrak{g}_2$& &$\mathfrak{f}_4$&
  &$\mathfrak{g}_2$&$\mathfrak{sp}(1)$& & &$\mathfrak{e}_8$& & &$\mathfrak{sp}(1)$ &$\mathfrak{g}_2$& &$\mathfrak{f}_4$&
  &$\mathfrak{g}_2$&$\mathfrak{sp}(1)$&  &\\
1 & 2&     2
                        &3&1&5&1&3&2&2&1&12&1&2&2&3&1&5&1&3&2&2 & $1^*$\\
  & &  & & & & & & & & &$\mid$ & & & & & & & & & &  &\\
  & &  & & & & & & & & & 1 & & & & & & & & & &  &\\
\hline
\end{tabular}} \, .
\end{equation}
Upon close inspection we find that  there are two instantons  on the $\mathfrak{e}_8$
divisor which yield the two-extra blow-ups with ${\mathrm I}_0$ fiber and no algebra indicated in \eqref{newE8n0}. 
In the result presented in \cite{Font:2016odl} the two instantons are not shown explicitly but implicitly understood from the fact that the self-intersection of the $\mathfrak e_8$-curve is given  by $-10$ \cite{Heckman:2015bfa}. 
Again, \eqref{newE8n0} should remind us not to forget about the two instantons at $\mathfrak e_8$ and,  
hence, we find  $52$ and $8$ for $h_{\mathrm R}$ and $r_{\mathrm R}$, respectively.
The same values of $h_{\mathrm R}$ and $r_{\mathrm R}$ are
also obtained for the resolution in the $\mathrm{Spin}(32)/\ZZ_2$ heterotic string which reads
\begin{equation}
\label{modE80}
\begin{tabular}{|cccccccc|}
\hline
 $\mathfrak{sp}(10)$ &  $\mathfrak{so}(24)$ &$\mathfrak{sp}(6)$& $\mathfrak{so}(16)$&
  $\mathfrak{sp}(2)$&$\mathfrak{so}(7)$& &
 $\mathfrak{so}(8) $ \\ 
1*&4&1&4&1&3&1&4 \\\hline
\end{tabular} \, .
\end{equation}
For the $\mathfrak{sp}(2) \oplus \mathfrak{so}(7)$ piece the matter hypermultiplets belong to 
$\frac12(\mathbf{fund}, \mathbf{spinor})$. 

\subsubsection{\texorpdfstring{$\kod{[IV^{\ast}-I_n]}$}{[IV^*-I_n]} Model and \texorpdfstring{$\mathrm{E}_6$}{E_6} Singularity}\label{ss:InE6}
%%%

Computing the moduli monodromy from the NU data for this model we obtain
\begin{equation}
\tau \rightarrow -\frac{1+\tau}{\tau}\, ,\quad \rho \rightarrow \rho + n - \frac{\beta^2}{\tau} \, ,\quad \beta \rightarrow
\frac{\beta}{\tau}
\, .
\end{equation}
Clearly  $\rho \to \rho + n$ when $\beta=0$ and the monodromy in $\tau$
is that of a $\kod{IV^{\ast}}$ type fiber. Thus, the model is conjectured
to describe $k=8+n$ instantons on an ${\mathrm E}_6$ singularity. 
Indeed, the resolutions in both heterotic strings produce the expected theories 
originally found in \cite{Aspinwall:1997ye}.
In the $E_8 \times E_8$ string the resolution for $n \ge 1$ gives
\begin{equation}
\label{resoInIVstar}
\begin{tabular}{|ccccccccc|}
\hline
 &&  $\mathfrak{sp}(1)$ &$\mathfrak{g}_2$&& $\mathfrak{f}_4$&
  &$\mathfrak{su}(3)$& \\
1 & 2&     2 &3&1&5&1&3&1\\\hline
\end{tabular}\, 
\begin{tabular}{|cccc|}
\hline
  $\mathfrak{e}_6$&  &$\mathfrak{su}(3)$& \\
6&1 &3&1\\\hline
\end{tabular}^{\, \otimes (n-1)} \,
\begin{tabular}{|cccccc|}
\hline
  $\mathfrak{f}_4$&&$\mathfrak{g}_2$&$\mathfrak{sp}(1)$& &\\
5 &1&3&2&2& $1^*$\\\hline
\end{tabular} \, .
\end{equation}
It is easy to verify that $h_{\mathrm R}=12 k   - 78$. 
Also $r_{\mathrm R}=6$ because matter just consists of
$\frac12(\mathbf{2},\mathbf{1}) \oplus \frac12(\mathbf{2},\mathbf{7})$ for each
$\mathfrak{sp}(1) \oplus \mathfrak{g}_2$. 
For $n=0$ we instead find
\begin{equation}
\label{resoe6n0}
\begin{tabular}{|ccccccccccc|}
\hline
 &&  $\mathfrak{sp}(1)$ &$\mathfrak{g}_2$&& $\mathfrak{f}_4$&
  &$\mathfrak{g}_2$& $\mathfrak{sp}(1)$ & &\\
1 & 2&     2 &3&1&4 &1&3&2&2& $1^*$\\\hline
\end{tabular} \, .
\end{equation}
The $\mathfrak{f}_4$ with self-intersection $-4$ comes with a hypermultiplet in the fundamental so again $r_{\mathrm R}=6$.

In the $\mathrm{Spin}(32)/\ZZ_2$  we obtain the resolution 
\begin{equation}
\label{modE6}
\begin{tabular}{|ccccc|}
\hline
$\mathfrak{sp}(k)$ & $\mathfrak{so}(4k\text{-}16)$ & $\mathfrak{sp}(3k\text{-}24)$ &
  $\mathfrak{su}(4k\text{-}32)$ & $\mathfrak{su}(2k\text{-}16)$\\
1*&4&1& 2 & 2\\
 \hline
 \end{tabular}  \, ,
\end{equation}
which is valid for $n \ge 1$, $k \ge 9$. The pattern 
conforms to the extended Dynkin diagram of $E_6$ but dropping the two outer nodes with
complex conjugate representations of the discrete subgroup $\Gamma_G$.
Matter includes $\frac12(\mathbf{fund}, \mathbf{fund})$ for adjacent $\mathfrak{sp}$-$\mathfrak{so}$
factors, but $(\mathbf{fund}, \mathbf{fund})$ for neighboring $\mathfrak{sp}$-$\mathfrak{su}$ and
$\mathfrak{su}$-$\mathfrak{su}$. For $\mathfrak{sp}(k)$ there are 16 extra fundamentals.
It can be checked that the values of $h_{\mathrm R}$ and $r_{\mathrm R}$ agree for both heterotic strings.
For $n=0$ the resolution gives 
\begin{equation}
\label{modE6n0}
\begin{tabular}{|ccccc|}
\hline
$\mathfrak{sp}(8)$ & $\mathfrak{so}(16)$ &  &
  & \\
 1*&4&1& 2 & 2\\
 \hline
 \end{tabular}  \, .
\end{equation}
There are four blow-ups but only one divisor supports a non-trivial algebra. In this case 
$r_{\mathrm R}$ is apparently 4 but the expected value 6 results from two extra neutral hypermultiplets, one from each curve with
self-intersection $-2$ and not attached to a non-Higgsable cluster \cite{Morrison:2012js}.

\subsubsection{\texorpdfstring{$[\kod I_n-\kod I_0^{\ast}]$}{[I_n-I_0^*]} Model and \texorpdfstring{$\mathrm{D}_4$}{D_4} Singularity}\label{ss:InI0s}

In the NU model $[\kod I_n-\kod I_p^{\ast}]$ the moduli monodromy is
\begin{equation}
\label{monodp4}
\tau \rightarrow \tau + p \, ,\quad \rho \rightarrow \rho + n 
\, ,\quad \beta \rightarrow - \beta \, .
\end{equation}
The monodromy in $\tau$ is that of a $\mathrm{I}_p^{\ast}$ type fiber. Given the monodromy in $\rho$ this model 
is then expected to describe $k = n + p + 6$ small instantons on a ${\mathrm D}_{p+4}$ singularity.
The resolutions for generic $p$ are discussed in Appendix~\ref{app:ADE}. Below we consider $p=0$. 

For $n\ge 1$, resolving the NU model in the $E_8 \times E_8$ string leads to
\begin{equation}
\label{singD4}
\begin{tabular}{|ccccc|}
\hline
 &&   $\mathfrak{sp}(1)$ & $\mathfrak{g}_2$& \\
1 &2&   2 &3  & 1 \\\hline
\end{tabular} \, 
\begin{tabular}{|cc|}
\hline 
$\mathfrak{so}(8)$& \\
4&1 \\\hline
\end{tabular}^{\, \oplus (n-1)}\, 
\begin{tabular}{|cccc|}
\hline
 $\mathfrak{g}_2$& $\mathfrak{sp}(1)$& &\\
3& 2 &2 & $1^*$\\\hline
\end{tabular} \, .
\end{equation}
For the quantity $h_{\mathrm R}$ we find $6 k - 28$, in agreement with \eqref{checkade}, and $r_{\mathrm R}=4$. 
When we set $n=0$, the resolution turns out to be
\begin{equation}
\label{singD40}
\begin{tabular}{|ccccccc|}
\hline
 &&  $\mathfrak{sp}(1)$ &$\mathfrak{g}_2$&$\mathfrak{sp}(1)$ & &\\
1 & 2&     2 &2&2&2 & $1^*$\\\hline
\end{tabular} \, .
\end{equation}
Thus $h_{\mathrm R}=10$. 
The matter content is $\frac12(\mathbf{2},\mathbf{1},\mathbf{1}) \oplus
\frac12(\mathbf{2},\mathbf{7},\mathbf{1}) \oplus 2 (\mathbf{1},\mathbf{7},\mathbf{1})
\oplus \frac12(\mathbf{1},\mathbf{7},\mathbf{2}) \oplus \frac12(\mathbf{1},\mathbf{1},\mathbf{2})$,
so $r_{\mathrm R}=4$ as it should.

Resolving the singularities coming from the $\mathrm{Spin}(32)/\ZZ_2$ heterotic setting, we derive
\begin{equation}
\label{modD4}
\begin{tabular}{|cc|}
\multicolumn{2}{c}{$n=0$}\\
\hline
$\mathfrak{sp}(6)$ &  $\mathfrak{so}(7)$\\ 
1*&1\\\hline
\end{tabular} \  ,  \hspace*{1cm} 
\begin{tabular}{|cc|}
\multicolumn{2}{c}{$n=1$}\\
\hline
$\mathfrak{sp}(7)$ &  $\mathfrak{so}(12)$\\ 
1*&1\\\hline
\end{tabular} \ , \hspace*{1cm} 
\begin{tabular}{|ccc|}
\multicolumn{3}{c}{$n\ge 2$}\\
\hline
& $\mathfrak{sp}(k\text{-}8)$ & \\
& 1 & \\
& $\mid$ & \\
$\mathfrak{sp}(k)$ & $\mathfrak{so}(4k\text{-}16)$ & $\mathfrak{sp}(k\text{-}8)$ \\ 
1*&4&1\\
& $\mid$ & \\
& 1 & \\
& $\mathfrak{sp}(k\text{-}8)$ & \\
\hline
\end{tabular}\, .
\end{equation}
The number of blow-ups is one for $n=0,1$, and four for $n\ge 2$. Thus, the values of $h_{\mathrm R}$
match for all $n$ in both heterotic strings. One can also check that $r_{\mathrm R}=4$ for all $n$.

\subsection{Non-Geometric Models and Dualities}
\label{sec:dual}

In the previous section we have seen that the explicit formulation of heterotic/F-theory
duality in terms of the map between genus-two and K3 fibrations
confirms the results expected from the moduli monodromies in models
corresponding to small instantons on ADE singularities.
We now turn to heterotic models with monodromies which
are non-geometric in all T-duality frames. This is the most interesting
situation, since there is no intuition about the nature of such degenerations  and it is
not even obvious whether they are allowed. 

A simple example of a non-geometric degeneration is the Namikawa-Ueno
$\kod{[III-III]}$ singularity which has monodromy
\begin{equation}
\label{ex33}
\tau \rightarrow \frac{\rho}{\beta^2-\rho\tau} \, ,\quad \rho
\rightarrow \frac{\tau}{\beta^2-\rho\tau} \, ,\quad \beta
\rightarrow -\frac{\beta}{\beta^2-\rho \tau} \, .
\end{equation} 
When $\beta = 0$ we obtain a ``double elliptic'' fibration with monodromy $\tau\to-1/\tau$, $\rho \to- 1/\rho $ when encircling the heterotic
degeneration. 
To study the model, we start with the equation of its defining hyperelliptic curve which is given by
\begin{equation}\label{eq:hyperell_III-III}
y^2 = x  (x - 1)  (x^{2} + t)  \left[(x-1)^2 + t\right]\, .
\end{equation}
Applying the resolution procedure we obtain the same six-dimensional theory derived in the $\kod{[I_0-I_0^{\ast}]}$ 
model, cf.~\eqref{modD4}, which is the theory of six small instantons on a ${\mathrm D}_4$ singularity.
In the $E_8 \times E_8$ string the same phenomenon occurs, namely the resulting theory is identical to \eqref{singD40}
\cite{Font:2016odl}. 

In the context of the $E_8 \times E_8$ heterotic string it was actually discovered that in several 
non-geometric models of type 2 in the NU list the dual CY admits a smooth resolution and, moreover, the emerging
low-energy physics is described by the theory of small instantons on ADE singularities \cite{Font:2016odl}. As an explanation it 
was argued that two NU models with the same resolution, such as $\kod{[III-III]}$ and $\kod{[I_0-I_0^{\ast}]}$, can be 
related by certain duality moves. Thus, the resolution of such dual models in the $\mathrm{Spin}(32)/\ZZ_2$ string 
is also expected to give theories equal to each other. One motivation behind this paper was precisely to address 
this issue. 

\begin{table}[t]\begin{center}
\renewcommand{\arraystretch}{1.5}
\begin{tabular}{|c|c|}
\hline
$\mu(c)$ & dual models   \\
\hline \hline
4& $\kod{[I_0-IV]}
  \, ,  \,\kod{[II-II]}$  \\ \hline 
5&$[\kod{IV}-\kod{I}_1] \, , \,\kod{[II-III]}$ \\ \hline 
6& $\kod{[I_0-I_0^{\ast}]}\, , \,
  \kod{[III-III]}\, , \,\kod{[IV-II]}$ \\ \hline 
 7& $[\kod{I}_0^{\ast}-\kod{I}_1]\, , \,\kod{[IV-III]}$ \\ \hline 
8& $\kod{[I_0-IV^{\ast}]}\, , \,
%%%
 \kod{[I_0^{\ast}-II]}$ \\ \hline 
9& $\kod{[I_0-III^{\ast}]} \,
  ,  \,\kod{[I_0^{\ast}-III]}$ \\ \hline 
10& $\kod{[I_0-II^{\ast}]} \, ,\,
%%%
 \kod{[I_0^{\ast}-IV]}$ \\ \hline 
11& $\kod{[II-III^{\ast}]} \, ,\,
  \kod{[IV^{\ast}-III]}$ 
\\ 
\hline 
\end{tabular}
\caption{Dual models: the NU degenerations in the same row give rise
  to the same LSTs after resolution of the dual F-theory model. }
  \label{tab:dual}\end{center}\end{table}

In the $\mathrm{Spin}(32)/\ZZ_2$ heterotic string
we find that dual models indeed appear as anticipated in \cite{Font:2016odl}.
A necessary condition is that the sum of the vanishing orders of the
discriminant for their two Kodaira components, or equivalently the
vanishing order  $\mu(c)$, is the same. In Table~\ref{tab:dual} we display
all the models satisfying this condition and admitting dual
smooth Calabi-Yau resolutions. For all the models in Table~\ref{tab:dual} we
explicitly performed the F-theory resolution. The results for models with small instantons on ADE 
singularities, cf.~Table~\ref{tab:adetype}, are presented
in Section~\ref{sec:ADE} and in Appendix~\ref{app:ADE}. The $\kod{[I_0-IV]}$  and $[\kod{IV}-\kod{I}_1]$ models 
correspond to $k = 4$ and $k = 5$ instantons on an ${\mathrm A}_2$ singularity, cf.~\eqref{modA2}.
We have verified that for all the degenerations in a row the same theory arises in both heterotic strings.

%%%
%%%
%%%
In the $E_8\times E_8$ setup the  $\kod{[II-IV^{\ast}]}$ model was included among the dual models
at $\mu(c)=10$ \cite{Font:2016odl}. However, we now find that the resolution of this model in the $E_8\times E_8$ 
string is actually given by 
\begin{equation}
\label{mod42e8}
\footnotesize{
\begin{tabular}{|ccccccccccccccccccccccc|}
\hline
  & &  & & & & & & & & & 2 & & & & & & & & & &  &\\
  & &  & & & & & & & & &$\mid$ & & & & & & & & & &  &\\
  & &  & & & & & & & & & 1 & & & & & & & & & &  &\\
 & &  & & & & & & & & &$\mid$ & & & & & & & & & &  &\\
 &  &$\mathfrak{sp}(1)$ &$\mathfrak{g}_2$& &$\mathfrak{f}_4$&
  &$\mathfrak{g}_2$&$\mathfrak{sp}(1)$& & &$\mathfrak{e}_8$& & &$\mathfrak{sp}(1)$ &$\mathfrak{g}_2$& &$\mathfrak{f}_4$&
  &$\mathfrak{g}_2$&$\mathfrak{sp}(1)$& &\\
1 & 2&     2
                        &3&1&5&1&3&2&2&1&12&1&2&2&3&1&5&1&3&2&2& $1^*$ \\
\hline
\end{tabular}} \ .
\end{equation}
%%%
%%%
%%%
%%%
%%%
In comparison with the resolution of $\kod{[I_0-II^{\ast}]}$ in \eqref{newE8n0} there is a difference in
the $\mathfrak{e}_8$ divisor. This time the two instantons lie on top of each other and, therefore, the required two-extra blow-ups, 
with ${\mathrm I}_0$ fiber and no algebra, have a different intersection structure. 
Besides, there is an extra neutral hypermultiplet from the $-2$ curve without a gauge algebra and 
not attached to a non-Higgsable cluster \cite{Morrison:2012js}. 
The two theories could be connected by RG flow as it occurs in analogous configurations \cite{Heckman:2013pva,Heckman:2015ola}.

%%%
%%%
%%%

%%%
%%%
%%%
%%%
%%%
%%%
%%%
%%%

In the  $\mathrm{Spin}(32)/\ZZ_2$ string the resolutions of $\kod{[II-IV^*]}$ and $\kod{[I_0-II^{\ast}]}$ do not coincide either.
The former reads
\begin{equation}
\label{mod24s}
\begin{tabular}{|cccccccc|}
\hline
 $\mathfrak{sp}(10)$ &  $\mathfrak{so}(24)$ &$\mathfrak{sp}(6)$& $\mathfrak{so}(16)$&
  $\mathfrak{sp}(2)$&$\mathfrak{so}(7)$& &
 $\mathfrak{so}(9) $ \\ 
1*&4&1&4&1&3&1& 4 \\\hline
\end{tabular} \, .
\end{equation} 
We observe that it differs from the resolution of $\kod{[I_0-II^{\ast}]}$ in \eqref{modE80} in the last algebra factor.
Now, the  $\mathfrak{so}(9)$ has a hypermultiplet in the fundamental which can break the symmetry to
 $\mathfrak{so}(8)$ so that the resolutions could match. 
To test whether the theories are really connected on a Higgs branch is an open question left for future investigations.
It is interesting to note that the $\kod{[II-IV^*]}$ and $\kod{[I_0-II^{\ast}]}$ theories are distinguished by the 
%%%
values of the intrinsic quantity $r_{\mathrm R}$, they are 9 and 8, respectively.

A second puzzling case is  the  $ \kod{[IV-IV]}$ model which was listed among the
duals at $\mu(c)=8$ in \cite{Font:2016odl}. Again we now find that the resolutions do not agree.
In the $\mathrm{Spin}(32)/\ZZ_2$ the resolved theory is 
\begin{equation}
\label{mod44}
\begin{tabular}{|ccccc|}
\hline
$\mathfrak{sp}(8)$ & $\mathfrak{so}(16)$ &  &
   & $\mathfrak{su}(2)$\\
1*&4&1& 2 & 2\\
 \hline
 \end{tabular}  \, ,
 \end{equation}
which could match the resolution of $\kod{[I_0-IV^{\ast}]}$  in \eqref{modE6n0} after the
$\mathfrak{su}(2)$ is higgsed away. 
In the $E_8 \times E_8$ heterotic string the resolution of $ \kod{[IV-IV]}$ gives
\begin{equation}
\label{res4s0}
\begin{tabular}{|ccccccccccc|}
\hline
 & &      &&&1 &&&&&\\
 & &      &&&$\mid$ &&&&&\\
 &&  $\mathfrak{sp}(1)$ &$\mathfrak{g}_2$&& $\mathfrak{f}_4$&
  &$\mathfrak{g}_2$& $\mathfrak{sp}(1)$ &&\\
1 & 2&     2 &3&1&5 &1&3&2&2& $1^*$\\\hline
\end{tabular} \, ,
\end{equation}
which differs from the resolution of $\kod{[I_0-IV^{\ast}]}$ in \eqref{resoe6n0} 
by having an additional blow-up attached to the  $\mathfrak{f}_4$ divisor which then has self-intersection $-5$
and no charged matter. The nature of a possible connection between the two theories is less clear.
It could be that when the additional tensor multiplet disappears, its degrees of freedom go into a ${\mathbf{26}}$
of $\mathfrak{f}_4$ plus three extra neutral hypermultiplets.
Again we notice that the $\kod{[IV-IV]}$ and $\kod{[I_0-IV^{\ast}]}$ theories have different values of
%%%
$r_{\mathrm R}$, cf.~\eqref{checkrank}, namely 9 and 6, respectively.
%%%

\section{The  \texorpdfstring{$\mathbf{\text{Spin}(32)/\ZZ_2}$}{Spin(32)/Z2} Catalog of T-fects}
\label{sec:catalog}

In this section, we summarize our findings about the Namikawa-Ueno models for which 
we could construct the dual CY resolution.   In both heterotic strings the resolvable models 
satisfy the criterion established in \cite{Font:2016odl}. A dual F-theory model with %%%
the coefficients $a$, $b$,
$c$, $d$, as in equation \eqref{eq:K328a}, has a resolution if and only if
$\mu(a)<4$ or $\mu(b)<6$ or $\mu(c)<10$ or $\mu(d)<12$ where $\mu$ is the vanishing order at $t=0$. 
Altogether there are 49 resolvable models out of the 120 entries in the NU 
classification.\footnote{The complete list of NU degenerations is reproduced in Appendix~D of \cite{Font:2016odl}.}
They are collected in the tables \ref{tab:e1p4p5},  \ref{tab:elliptic2} and  \ref{tab:parabolic3}, where
NU models $[\mathrm{K}_1-\mathrm{K}_2-0]$ are denoted
$[\mathrm{K}_1-\mathrm{K}_2]$.
The resulting theories in the  $E_8 \times E_8$ string were reported in \cite{Font:2016odl}. 
Here we complete the study of T-fects by working out the resolutions in the $\mathrm{Spin}(32)/\ZZ_2$ heterotic string. 

To present the results, we will consider separately the five different types in the NU classification.
Besides the local equation of the genus-two degeneration given by a sextic over $t \in \CC$, NU also provide the $Sp(4,\ZZ)$ 
monodromy around the singularity at $t=0$. A model is called elliptic or parabolic if the monodromy is of finite or infinite order.
Each degeneration is further characterized by the type of singular fiber or equivalently by the modulus point, which 
is a fixed point of the monodromy and belongs to the compactification of  ${\mathbb H}_2/Sp(4,\ZZ)$ \cite{Namikawa:1973yq}. 
For instance, in the elliptic type 2 models the modulus point has the Wilson line $\beta$ identically zero
and the singular fiber consists of two elliptic curves meeting at one point.

As in the examples discussed in the previous section, in each case the analysis begins with the local
equation from which we determine the dual F-theory background. We then apply the resolution procedure
described in Section~\ref{ss:toric}. For every model admitting a resolution we compute the self-intersection numbers
of the $n_T$ blow-up divisors and derive the gauge algebra supported at each of them.
Taking into account the $\mathfrak{sp}(k)$ factor supported at $t=0$, the resulting theory is encoded in a tree-like diagram 
with $n_T+1$ nodes. The matter content can also be derived and shown to be compatible with anomaly cancellation.
The pure gauge anomaly takes the form \eqref{gsfact}. In all models $\eta^{\alpha\beta}$ is positive semi-definite with one
null eigenvalue.
We have also verified agreement in both heterotic strings of the quantities $h_{\mathrm R}$ and
$r_{\mathrm R}$ defined in \eqref{checkdef} and \eqref{checkrank}.

In the models with $\mu(d) > \mu(c)$ matter includes hypermultiplets arising from intersections with the global 
$\mathrm{Spin}(28) \times SU(2)/\ZZ_2$ as we explained before. In each case the precise blow-up divisor that intersects the 
$\mathrm{I}_2$ curve can be determined from the toric data of the resolution.  
%%%
%%%
The results can be confirmed by anomaly cancellation. Below we will give some representative examples.

\subsection{Elliptic Type 1}

The elliptic type 1 NU degenerations are distinguished by a monodromy action
that mixes the three moduli. Even though the corresponding
heterotic models lack a geometric interpretation, the dual F-theory
resolutions are analogous to those discussed in Section~\ref{sec:ADE}.
In Table~\ref{tab:e1p4p5} we show the models whose F-theory duals admit 
a smooth CY resolution. The non-trivial nonequivalent resolutions are displayed below.
Models $\kod{[V]}$ and $\kod{[VII]}$ have the same resolution. 
 
\begin{table}[h!]\begin{center}
\renewcommand{\arraystretch}{1.5}
\begin{tabular}{|M{1cm}|c|c|c|c|c|}
\hline
class &  NU model & $\mu(a)$ & $\mu(b)$ & $\mu(c)$ & $\mu(d)$  \\
\hline\hline
%%%
%%%
\multirow{5}{*}{\parbox{1cm}{\tiny{Elliptic type 1}}} &   $\kod{[I_{0-0-0}]}$  &0& 0&0&0 \\ \cline{2-6} 
& $\kod{[V]}$  &2& 3&5&6 \\ \cline{2-6} 
 & $\kod{[VII]}$ &2& 3&5&6 \\ \cline{2-6} 
& $\kod{[VIII-1]} $&$\infty$& $\infty$&4&$\infty$ \\ \cline{2-6} 
& $\kod{[IX-1]}$ &$\infty$& $\infty$&8&$\infty$  \\
\hline
 %%%
 %%%
\hline
\multirow{3}{*}{\parbox{1cm}{\tiny{Parabolic type 4}}} & $[\kod{I}_{n-p-0}]$  &0&0&$n+p$&$n+p$ \\ \cline{2-6} 
 & $[\kod{I}_n-\kod I_p^{\ast}]$  &2&3&$6+n+p$&$6+n+p$ \\ \cline{2-6} 
& $[\kod{II}_{n-p}]$  &2&3&$5+n+p$&$6+n+p$ \\ \cline{2-6} 
%%%
%%%
 \hline\hline
\multirow{2}{*}{\parbox{1cm}{\tiny{Parabolic type 5}}} & $[\kod{I}_{n-p-q}]$  &0&0&$n+p+q$&$n+p+q$ \\ \cline{2-6} 
& $[\kod{II}_{n-p}]$  {\scriptsize$p=2k+l$, $ l=0,1$ }&2&3&$5+l+2k+n$&$6+l+2k+n$ \\ 
\hline
\end{tabular}
\caption{Elliptic type 1, parabolic type 4 and parabolic type 5 resolvable models.}
  \label{tab:e1p4p5}\end{center}\end{table}

\subsubsection*{\texorpdfstring{$\kod{[V]}$}{[V]} Model} 

\begin{equation}
\label{modV}
\begin{tabular}{|cc|}
\hline
$\mathfrak{sp}(5)$ &  $\mathfrak{so}(7)$\\ 
1*&1\\\hline
\end{tabular}  \, .
\end{equation}

\subsubsection*{\texorpdfstring{$\kod{[VIII-1]}$}{[VIII-1]} Model}

\begin{equation}
\label{modVIII}
\begin{tabular}{|cc|}
\hline
$\mathfrak{sp}(4)$ &  $\mathfrak{su}(2)$\\ 
1*&1\\\hline
\end{tabular}  \, .
\end{equation}

\subsubsection*{\texorpdfstring{$\kod{[IX-1]}$}{[IX-1]} Model}
\begin{equation}
\label{modIX}
\begin{tabular}{|ccccccc|}
\hline
$\mathfrak{sp}(8)$ &  $\mathfrak{so}(20)$ &$\mathfrak{sp}(4)$& $\mathfrak{so}(12)$& &
  $\mathfrak{su}(2)$&$\mathfrak{so}(7)$ \\ 
1*&4&1&4&1&2&3 \\\hline
\end{tabular} \, .
\end{equation}

\subsection{Elliptic Type 2}

The 20 models that can be resolved are listed in
Table~\ref{tab:elliptic2}. The resolutions of 
models of type $\kod{[I_0-K_2]}$, corresponding to configurations of
$k=\mu(c)$ pointlike instantons on the $\kod{K_2}$ singularity, are reviewed
in Section~\ref{sec:ADE} and Appendix~\ref{app:ADE}.
Other models are non-geometric because
their monodromy involves a non-trivial action on the torus volume. 
However, as discussed in Section~\ref{sec:dual}, many of these
models have the same resolutions as the geometric ones. An intriguing model in this class is
$\kod{[II-III^{\ast}]}$, dual to $ \kod{[IV^{\ast}-III]}$, whose resolution involves the exceptional algebra $\mathfrak{e}_7$ as shown below.

\subsubsection*{\texorpdfstring{$\kod{[II-III^{\ast}]}$}{[II-III^*]} Model}

\begin{equation}
\label{mod23s}
\begin{tabular}{|ccccccccccccccc|}
\hline
$\mathfrak{sp}_{11}$ & $\mathfrak{so}_{28}$  &  $\mathfrak{sp}_9$ & $\mathfrak{so}_{24}$ & $\mathfrak{sp}_7$ &
$\mathfrak{so}_{20}$ & $\mathfrak{sp}_5$ & $\mathfrak{so}_{16}$ & $\mathfrak{sp}_3$ & $\mathfrak{so}_{12}$ &
$\mathfrak{sp}_1$ & $\mathfrak{so}_7$ & $\mathfrak{su}_2$ & & $\mathfrak{e}_7$ \\
1* & 4 & 1 & 4 & 1 & 4 & 1 & 4 & 1 & 4 & 1 & 3 & 2 & 1 & 8 \\
\hline
\end{tabular} \, .
\end{equation}

\begin{table}[t]\begin{center}
\renewcommand{\arraystretch}{1.5}
\begin{tabular}{|c|c|c|c|c||c|c|c|c|c|}
\hline
NU model & $\mu(a)$ & $\mu(b)$ & $\mu(c)$ & $\mu(d)$&NU model & $\mu(a)$ & $\mu(b)$ & $\mu(c)$ & $\mu(d)$  \\
\hline \hline
$\kod{[I_0-I_0]}$  &0&0&0&0 &$\kod{[II-IV]}$ &3& 3&6&6 \\ \hline 
  $\kod{[I_0-II]}$ &1& 1&2&2 &$\kod{[I_0^{\ast}-II]}$ &3& 4&8&8 \\ \hline 
 $\kod{[I_0-III]} $&1& 2&3&3 &$\kod{[II-IV^{\ast}]}$ &5& 5&10&10 \\ \hline 
$\kod{[I_0-IV]}$ &2& 2&4&4  &$\kod{[II-III^{\ast}]}$ &4& 7&11&11 \\\hline 
$\kod{[I_0-I_0^{\ast}]}$ &2& 3&6&6  &$\kod{[III-III]}$ &2& 4&6&6 \\\hline 
$\kod{[I_0-IV^{\ast}]}$ &4& 4&8&8  &$\kod{[IV-III]}$ &3& 4&7&7 \\\hline 
$\kod{[I_0-III^{\ast}]}$ &3& 6&9&9  &$\kod{[I_0^{\ast}-III]}$ &3& 5&9&9 \\\hline 
$\kod{[I_0-II^{\ast}]}$ &5& 5&10&10 &$\kod{[IV^{\ast}-III]}$ &5& 6&11&11  \\\hline 
$\kod{[II-II]}$ &2& 2&4&4 &$\kod{[IV-IV]}$ &4& 4&8&8  \\\hline 
$\kod{[II-III]}$ &2&3&5&5 &$\kod{[I_0^{\ast}-IV]}$ &4& 5&10&10 \\\hline 
%%%
\end{tabular}
\caption{Elliptic type 2 resolvable models.}  \label{tab:elliptic2}\end{center}\end{table}

\subsection{Parabolic Type 3}

Table \ref{tab:parabolic3} contains the 19 parabolic type 3 degenerations whose dual F-theory CY can 
be resolved. They admit a resolution for all $n$. The
models of type $[\kod{I}_n-\kod K_2]$ or $[\kod{K}_1-\kod{I}_n]$ again describe
$k=\mu(c)$ pointlike instantons on the $\kod{K}_i$ singularity and their
resolution is shown in Section~\ref{sec:ADE} or Appendix~\ref{app:ADE}. The resolution for 
$[\kod{II}_{n-0}]$ and other non-trivial examples will be given below.
In this class we also discover dual models. Specifically, starting with the fifth row in Table~\ref{tab:parabolic3}, the
models in the same row have the same resolution.

\subsubsection*{\texorpdfstring{$[\mathrm{II}_{n-0}]$}{[II_{n-0}]} Model}

\begin{equation}
\label{modIIn0}
\begin{tabular}{|ccc|}
\multicolumn{3}{l}{$n=1$}\\
\hline
$\mathfrak{sp}(6)$ &  $\mathfrak{so}(12)$ & \\ 
1*&2&1\\\hline
\end{tabular} \ , \hspace*{1cm} 
\begin{tabular}{|ccc|}
\multicolumn{3}{l}{$n\ge 2$}\\
\hline
& $\mathfrak{sp}(n\text{-}2)$ & \\
& 1 & \\
& $\mid$ & \\
$\mathfrak{sp}(n\text{+}5)$ & $\mathfrak{so}(4n\text{+}8)$ & $\mathfrak{sp}(n\text{-}1)$ \\ 
1*&4&1\\
& $\mid$ & \\
& 1 & \\
& $\mathfrak{sp}(n\text{-}2)$ & \\
\hline
\end{tabular}\, .
\end{equation}

\begin{table}[h!]\begin{center}
\renewcommand{\arraystretch}{1.5}
\begin{tabular}{|c|c|c|c|c||c|c|c|c|c|}
\hline
NU model & $\mu(a)$ & $\mu(b)$ & $\mu(c)$ & $\mu(d)$&NU model & $\mu(a)$ & $\mu(b)$ & $\mu(c)$ & $\mu(d)$  \\
\hline \hline
$[\kod I_{n-0-0}]$  &0&0&$n$&$n$ &
 $[\kod{II}-\kod I_n] $&$1+n$& 1&$2+n$&$2+n$ \\ \hline 
$[\kod{III}-\kod I_n]$ &1& $2+n$&$3+n$&$3+n$  &
$[\kod{III}-\kod{II}_n]$ &1& $2+n$&$3+n$&$4+n$ \\ \hline 
$[\kod{IV}-\kod{I}_n]$ &$2+n$& 2&$4+n$&$4+n$  &
$[\kod{IV}-\kod{II}_n]$ &$2+n$& 2&$4+n$&$5+n$ \\ \hline 
$[\kod{II}_{n-0}]$ &2& 3&$5+n$&$6+n$ &
& & & & \\ \hline 
$[\kod I_n- \kod I_0^{\ast}]$ &2& 3&$6+n$&$6+n$  &
$[\kod{I}_0-\kod I_n^{\ast}]$ &2&3&$6+n$&$6+n$ \\ \hline
$[\kod{IV}^{\ast}-\kod I_n]$ &$4+n$& 4&$8+n$&$8+n$ &
$[\kod{II}-\kod I_n^{\ast}]$ &3&4&$8+n$&$8+n$ \\ \hline
$[\kod{III}^{\ast}-\kod I_n]$ &3& $6+n$&$9+n$&$9+n$ &
$[\kod{III}-\kod I_n^{\ast}]$ &3& 5&$9+n$&$9+n$ \\ \hline 
$[\kod{II}^{\ast}-\kod I_n]$ &$5+n$& 5&$10+n$&$10+n$ &
$[\kod{IV}-\kod I_n^{\ast}]$ &4& 5&$10+n$&$10+n$ \\ \hline 
$[\kod{IV}^{\ast}-\kod{II}_n]$ &$3+n$& 4&$7+n$&$9+n$ &
$[\kod{II}-\kod{II}_n^{\ast}]$ &$3+n$& 4&$7+n$&$9+3n$  \\ \hline 
$[\kod{III}^{\ast}-\kod{II}_n]$ &3& $5+n$&$8+n$&$11+n$  &
$[\kod{III}-\kod{II}_n^{\ast}]$ &3& $5+n$&$8+n$&$10+2n$ \\ \hline 
%%%
\end{tabular}
\caption{Parabolic type 3 resolvable models.}
  \label{tab:parabolic3}\end{center}\end{table}

\subsubsection*{\texorpdfstring{$[\mathrm{IV^*}-\mathrm{II}_n]$}{[IV^*-II_n]} Model, $n\ge 1$}
\begin{equation}
\label{modIVsIIn}
\begin{tabular}{|ccccc|}
\hline
$\mathfrak{sp}(n\text{+}7)$ &  $\mathfrak{so}(4n\text{+}16)$ &$\mathfrak{sp}(3n\text{+}1)$& 
$\mathfrak{su}(4n\text{+}2)$& $\mathfrak{su}(2n\text{+}2)$ \\ 
1*&4&1&2&2 \\\hline
\end{tabular} \, .
\end{equation}

\subsubsection*{\texorpdfstring{$[\mathrm{III}-\mathrm{II}_n^*]$}{[III-II^*_n]} Model,  $n\ge 1$}

\begin{equation}
\label{mod32ns}
\begin{tabular}{|ccccccc|}
\hline
  &   &   & $\mathfrak{sp}(2n\text{-}1)$ &   &    &   \\
    &   &   & 1  &   &    &   \\
    &   &   & $\mid$ &   &    &   \\  
$\mathfrak{sp}(n\text{+}8)$ &  $\mathfrak{so}(4n\text{+}20)$ &$\mathfrak{sp}(3n\text{+}4)$& 
$\mathfrak{so}(8n\text{+}12)$& $\mathfrak{sp}(3n\text{+}1)$ & $\mathfrak{so}(4n\text{+}8)$ &
$\mathfrak{sp}(n\text{-}1)$ \\ 
1* & 4 & 1 & 4 & 1 & 4 & 1 \\\hline
\end{tabular} \, .
\end{equation}

\subsubsection*{\texorpdfstring{$[\mathrm{III}-\mathrm{II}_n]$}{[III-II_n]} Model}

\begin{equation}
\label{mod32n}
\begin{tabular}{|cc|}
\hline
$\mathfrak{sp}(n\text{+}3)$ &  $\mathfrak{sp}(n)$\\ 
1*&1\\\hline
\end{tabular}  \, .
\end{equation}

\subsubsection*{\texorpdfstring{$[\mathrm{IV}-\mathrm{II}_n]$}{[IV-II_n]} Model}

\begin{equation}
\label{mod42n}
\begin{tabular}{|cc|}
\hline
$\mathfrak{sp}(n\text{+}4)$ &  $\mathfrak{su}(2n\text{+}2)$\\ 
1*&1\\\hline
\end{tabular}  \, .
\end{equation}

\subsection{Parabolic Type 4}\label{sec:parabolic-type-4}

Only the three models shown in Table~\ref{tab:e1p4p5} admit a dual smooth resolution. The explicit resolutions of 
$[\kod{I}_{n-p-0}]$ and $[\kod{I}_n-\kod{I}_p^{\ast}]$ are given in Appendix~\ref{app:ADE}. 
The $[\kod{II}_{n-p}]$ resolution is shown below. In this case, it is instructive to look at  the intersections with the global $SU(2)$.
The toric analysis shows that the $\mathrm{I}_2$ curve intersects the divisor supporting $\mathfrak{sp}(m\text{-}1)$. Thus, there are
two extra fundamentals of $\mathfrak{sp}(m\text{-}1)$. Taking into account the matter from the intersection with $\mathfrak{so}(4m\text{+}8)$ 
gives the $(2m+6)$ fundamentals needed for anomaly cancellation.

\subsubsection*{\texorpdfstring{$[\mathrm{II}_{n-p}]$}{[II_{n-p}]} Model, \texorpdfstring{$m=n+p$, $\ell=6+n-p$}{m=n+p, l=6+n-p}}

\begin{equation}
\footnotesize{
\label{modIInpeven}
\begin{tabular}{|cccccccccc|}
\multicolumn{3}{l}{\normalsize{$p$ even, $n\ge p\text{+}2$, $n_T=p+4$}}\\
\hline
 & $\mathfrak{sp}(m\text{-}1)$ & & & & & & & $\mathfrak{sp}(\ell\text{-}8)$ &  \\
 & $\mid$ & & & & & & & $\mid$ &  \\
 & 1 & & & & & & & 1 &  \\
$\mathfrak{sp}(m\text{+}5)$ &  $\mathfrak{so}(4m\text{+}8)$ &$\mathfrak{sp}(2m\text{-}4)$& $\mathfrak{so}(4m\text{-}8)$&
  $\mathfrak{sp}(2m\text{-}12)$& $\cdots $ & $\mathfrak{so}(4\ell)$&$\mathfrak{sp}(2\ell\text{-}8)$&
 $\mathfrak{so}(4\ell\text{-}16) $ & $\mathfrak{sp}(\ell\text{-}8)$ \\ 
1*&4&1&4&1& $\cdots$ & 4&1&4 & 1\\\hline
\end{tabular} \, .
}
\end{equation}

%%%

\begin{equation}
\footnotesize{
\label{modIInpodd}
\begin{tabular}{|ccccccccc|}
\multicolumn{3}{l}{\normalsize{$p$ odd, $n\ge p$, $n_T=p+3$}}\\
\hline
 & $\mathfrak{sp}(m\text{-}1)$ & & & & & & &   \\
 & $\mid$ & & & & & & &   \\
 & 1 & & & & & & &   \\
$\mathfrak{sp}(m\text{+}5)$ &  $\mathfrak{so}(4m\text{+}8)$ &$\mathfrak{sp}(2m\text{-}4)$& $\mathfrak{so}(4m\text{-}8)$&
  $\mathfrak{sp}(2m\text{-}12)$& $\cdots $ & $\mathfrak{so}(4\ell\text{-}8)$&$\mathfrak{sp}(2\ell\text{-}12)$&
 $\mathfrak{su}(2\ell\text{-}12) $  \\ 
1*&4&1&4&1& $\cdots$ & 4&1&2\\\hline
\end{tabular} \, .
}
\end{equation}

\subsection{Parabolic Type 5}

{}From the 6 NU degenerations of type 5 only the two in Table~\ref{tab:e1p4p5} have a dual F-theory CY that can be
resolved.   
We find that the configuration derived from $[\kod{I}_{n-p-q}]$ has some analogies with the theory of small instantons 
on ${\mathrm A}$-type singularities, described by the model $[\kod{I}_{n-p-0}]$ considered in Appendix~\ref{app:ADE}. Despite sharing the name the parabolic type 5 and type 4 $[\kod{II}_{n-p}]$ models are not the same and 
their resolutions are different.

\subsubsection*{\texorpdfstring{$[\kod{I}_{n-p-q}]$}{[I_{n-p-q}]} Model}

We assume for simplicity that $n > p > q$. The result is actually completely symmetric under permutations of $(n, p, q)$. It is
convenient to introduce the auxiliary quantities $m=p+q$, $k=n+m$, and $\ell=k-2m$. The number of blow-ups is $n_T=[m/2]$.
Concerning the gauge algebra, next to $\mathfrak{sp}(k)$, starting  with $\mathfrak{su}(2k\text{-}6)$, there is a chain of $q$ $\mathfrak{su}$ factors,
in which the rank descends in units of six. When $m$ is odd and $p>q+1$ there follows a second chain  of $\mathfrak{su}$ factors
with rank descending by eight.  When $m$ is even and $p>q+2$ there is also such a second  $\mathfrak{su}$ chain 
with rank jumping in units of eight, plus an $\mathfrak{sp}(\ell+q)$ algebra at the end. 
These patterns are  represented below.

\begin{equation}
\footnotesize{
\label{modInpqeven}
\begin{tabular}{|cccccccccc|}
\multicolumn{3}{l}{$m\  \text{even}, n_T=m/2$}\\
\hline
$\mathfrak{sp}(k)$ &  $\mathfrak{su}(2k\text{-}6)$ &$\mathfrak{su}(2k\text{-}12)$& $\cdots$&
  $\mathfrak{su}(2k\text{-}6q)$& $\mathfrak{su}(2k\text{-}6q\text{-}8)$&$\mathfrak{su}(2k\text{-}6q\text{-}16)$&
$\cdots$ &  $\mathfrak{su}(2\ell\text{+}2q\text{+}8) $ & $\mathfrak{sp}(\ell\text{+}q)$ \\ 
1*&2&2&$\cdots$ &2& 2 & 2 &$\cdots$ &2 & 1\\\hline
\end{tabular} \, .
}
\end{equation}

\begin{equation}
\footnotesize{
\label{modInpqodd}
\begin{tabular}{|ccccccccc|}
\multicolumn{3}{l}{$m\  \text{odd}, n_T=(m-1)/2$}\\
\hline
$\mathfrak{sp}(k)$ &  $\mathfrak{su}(2k\text{-}6)$ &$\mathfrak{su}(2k\text{-}12)$& $\cdots$&
  $\mathfrak{su}(2k\text{-}6q)$& $\mathfrak{su}(2k\text{-}6q\text{-}8)$&$\mathfrak{su}(2k\text{-}6q\text{-}16)$&
$\cdots$ &  $\mathfrak{su}(2\ell\text{+}2q\text{+}4) $  \\ 
1*&2&2&$\cdots$ &2& 2 & 2 &$\cdots$ & 1\\\hline
\end{tabular} \, .
}
\end{equation}

\subsubsection*{\texorpdfstring{$[\mathrm{II}_{n-p}]$}{[II_{n-p}]} Model, \texorpdfstring{$m=n+p$, $\ell=n-\left[p/2\right]$}{m=n+p, l=n-[p/2]}}

Also for this model it is instructive to study the intersections with the global $SU(2)$. We find that the
$\mathrm{I}_2$ curve intersects the divisor supporting $\mathfrak{sp}(\ell\text{-}1)$ for $p$ even, and 
 $\mathfrak{su}(2\ell) $ for $p$ odd. In both cases there are two additional fundamentals which are
 required to precisely cancel the anomaly.

\begin{equation}
\footnotesize{
\label{modIInp5even}
\begin{tabular}{|cccccccccc|}
\multicolumn{3}{l}{\normalsize{$p$ even, $n\ge p\text{+}1$,  $n_T=p+4$}}\\
\hline
 & $\mathfrak{sp}(m\text{-}2)$ & & & & & & & $\mathfrak{sp}(\ell\text{-}2)$ &  \\
 & $\mid$ & & & & & & & $\mid$ &  \\
 & 1 & & & & & & & 1 &  \\
$\mathfrak{sp}(m\text{+}5)$ &  $\mathfrak{so}(4m\text{+}8)$ &$\mathfrak{sp}(2m\text{-}3)$& $\mathfrak{so}(4m\text{-}4)$&
  $\mathfrak{sp}(2m\text{-}9)$& $\cdots $ & $\mathfrak{so}(4\ell\text{+}20)$&$\mathfrak{sp}(2\ell\text{+}3)$&
 $\mathfrak{so}(4\ell\text{+}8) $ & $\mathfrak{sp}(\ell\text{-}1)$ \\ 
1*&4&1&4&1& $\cdots$ & 4&1&4 & 1\\\hline
\end{tabular} \, .
}
\end{equation}

%%%

\begin{equation}
\footnotesize{
\label{modIInp5odd}
\begin{tabular}{|ccccccccc|}
\multicolumn{3}{l}{\normalsize{$p$ odd, $n\ge p$,  $n_T=p+3$}}\\
\hline
 & $\mathfrak{sp}(m\text{-}2)$ & & & & & & &  \\
 & $\mid$ & & & & & & &   \\
 & 1 & & & & & & &   \\
$\mathfrak{sp}(m\text{+}5)$ &  $\mathfrak{so}(4m\text{+}8)$ &$\mathfrak{sp}(2m\text{-}3)$& $\mathfrak{so}(4m\text{-}4)$&
  $\mathfrak{sp}(2m\text{-}9)$& $\cdots $ & $\mathfrak{so}(4\ell\text{+}12)$&$\mathfrak{sp}(2\ell\text{-}1)$&
 $\mathfrak{su}(2\ell) $  \\ 
1*&4&1&4&1& $\cdots$ & 4&1& 2\\\hline
\end{tabular} \, .
}
\end{equation}

\section{Final Comments}

In this article we have further studied six-dimensional ${\mathcal N}=(1,0)$ non-geometric heterotic vacua described locally as 
$T^2$ fibrations over a complex one-dimensional base. More precisely, the moduli of the heterotic string 
compactified on $T^2$ are allowed to vary over the base and to
transform under monodromies in the duality group around points on the base. %%%
We have considered configurations with the gauge group broken by a background with $SU(2)$ structure in which
case the moduli are a single complex Wilson line modulus 
plus the complex structure and the complexified K\"ahler modulus of $T^2$.
The heterotic duality group is then $O(2,3,\ZZ)$ and there is a map between the heterotic moduli space and the moduli space
of genus-two curves. 
Thus, the non-geometric heterotic vacua can be defined equivalently as fibrations of a genus-two Riemann 
surface over the base. 
Even though genus-two fibrations have monodromies only in $Sp(4,\ZZ) \subset O(2,3,\ZZ)$, they are
specially tractable because their degenerations over a complex one-dimensional base have been classified by 
Namikawa and Ueno (NU) who also provided the corresponding moduli monodromies \cite{Namikawa:1973yq}.
The  $Sp(4,\ZZ)$ duality transformation around a degeneration signals the presence of defects.
In the case of the $E_8 \times E_8$ heterotic string, the six-dimensional theories living on the defects associated to the degenerations in the NU list were examined in 
 \cite{Font:2016odl}. In this work we focused on the
$\mathrm{Spin}(32)/\ZZ_2$ heterotic string.

Our approach relies on the heterotic/F-theory duality which relates the heterotic string compactified on $T^2$ and 
F-theory compactified on an elliptically fibered K3 surface.
In our setup, with one Wilson line breaking the gauge group to  
$\mathrm{Spin}(28)\times SU(2)/\ZZ_2$,  
the explicit map from the heterotic to the dual K3 moduli can be written in terms of Siegel 
modular forms of the genus-two curve encoding the heterotic moduli \cite{Clingher:2146c,Clingher:3503c,Malmendier:2014uka}.
Therefore, in F-theory, the non-geometric heterotic vacua described as genus-two fibrations over a base are realized as
specific K3 fibrations over the same base. We also know that when the base is complex one-dimensional
the total space of the F-theory fibration must be a Calabi-Yau threefold to preserve supersymmetry.
Therefore, the strategy is to use the well-defined geometric formalism of F-theory to analyze the degenerations of the fiber along the
base. This way, we have been able to determine the six-dimensional ${\mathcal N}=(1,0)$ theories living on defects
associated to genus-two degenerations in the NU classification.

The NU degenerations are given in terms  of fibrations of hyperelliptic curves defined by sextics with a singularity at a canonical
point on the base. For every such sextic we obtained the dual F-theory K3 which necessarily degenerates at the same point.
We then attempted to resolve the singularity in the F-theory picture by applying the toric inspired procedure explained in section \ref{sec:resolution}.
{}From the 120 types in the NU classification we found that only 49 lead to F-theory duals admitting a resolution by $n_T$ base blow-ups with $n_T$ finite. 

Introducing base blow-ups amounts to moving onto the tensor branch of the 6d \mbox{${\mathcal N}=(1,0)$} theory
living on the defect by turning on vevs of scalars in $n_T$ tensor multiplets.  For the resolvable models,
we obtained the theory emerging in the IR at a generic point on the tensor branch. 
Besides the number of tensor multiplets, the theory is characterized by matter hypermultiplets and
vector multiplets of a gauge algebra composed by factors supported at the blow-up divisors. 
In the  $\mathrm{Spin}(32)/\ZZ_2$ heterotic string there always appears a gauge factor that is 
not supported at a blow-up divisor, namely an ${\mathfrak{sp}}(k)$, where $k$ depends on the particular NU singularity.
Moreover, we find that there is charged matter due to intersections of divisors supporting gauge
algebras with the locus of the unbroken $\mathrm{Spin}(28)\times SU(2)/\ZZ_2$.

In the end, the theory resulting from a resolution is captured by a quiver diagram with $n_T+1$ nodes which encodes 
the full gauge algebra and matter content. In particular, the adjacency matrix that determines the pure gauge anomaly
can be read off from the diagram. In all resolvable models this matrix proves to be positive semi-definite
with one null eigenvalue. This implies in particular that the pure gauge anomaly can be cancelled by the Green-Schwarz-Sagnotti
mechanism  involving precisely $n_T$ tensor multiplets. From the existence of one null eigenvalue
it also follows that a linear combination of gauge couplings is independent of the scalars in the tensor 
multiplets and therefore it defines a mass scale. In turn, the presence of
such a scale indicates that the UV completion of the theories are little string theories (LSTs)
\cite{Seiberg:1997zk}. The resulting theories actually fall into recent classifications of LSTs \cite{Bhardwaj:2015xxa, Bhardwaj:2015oru}. 
Furthermore, deleting the node corresponding to ${\mathfrak{sp}}(k)$, which becomes a flavor symmetry, 
gives the tensor branch realization of 6d SCFTs embedded in the LSTs \cite{Bhardwaj:2015oru}. 

The resolvable NU degenerations are the same in the $\mathrm{Spin}(32)/\ZZ_2$ and
the $E_8 \times E_8$ heterotic strings, as expected since the two are related by T-duality upon circle
compactification. The T-duality manifests itself in a double fibration structure of the dual F-theory K3, which
has been claimed to be a necessary condition to realize LSTs in F-theory constructions \cite{Bhardwaj:2015oru}. 
To probe the T-duality we compared the LSTs emerging from the NU resolvable degenerations in both heterotic strings.
In all 49 cases we found that there are two intrinsic quantities that match. 
%%%

In the class of NU degenerations in which the moduli monodromies imply that the associated defects correspond to small
instantons on ADE singularities, we obtained theories that completely reproduce known 
results \cite{Aspinwall:1997ye, Blum:1997mm}. In many other NU degenerations with
non-geometric monodromies, we deduced novel theories that provide concrete examples of LSTs and embedded SCFTs.
They can serve as testing grounds to study properties of LSTs and SCFTs along the lines of recent investigations
\cite{Heckman:2015ola, Cordova:2015fha, Heckman:2015axa, Heckman:2016ssk, Morrison:2016djb, Mekareeya:2016yal, Mekareeya:2017jgc,
Mekareeya:2017sqh, Apruzzi:2017iqe}.

Also in  the $\mathrm{Spin}(32)/\ZZ_2$ heterotic string, we find that in several cases a resolvable NU degeneration with non-geometric 
moduli monodromies gives rise to the same theory obtained from another NU degeneration describing small instantons on a particular ADE 
singularity. This occurs when the monodromies of the two NU models are related by a chain of duality moves, as observed
previously in  the $E_8\times E_8$ heterotic string \cite{Font:2016odl}.
However, in both strings we have detected a couple of examples in which two resolvable degenerations lead to different theories
despite being related by duality moves. Nonetheless, it is plausible that the two theories are connected by RG flow.
It would be interesting to address this problem in more detail in the future.
%%%

\vspace{1cm}
\textbf{Acknowledgments:} 
We are indebted to I\~naki Garc\'{\i}a-Extebarria, Dieter L\"ust, and Stefano Massai for  many 
enlightening discussions and collaboration in the  early stage of this project.
We are also grateful to Stefan Theisen for helpful remarks. 
The research of C.M.~is supported by the Munich Excellence Cluster for Fundamental Physics ``Origin and the Structure of the Universe.'' 
A.F.~thanks the Institut Henri Poincar\'e, the Max-Planck-Institut f\"ur Gravitationsphysik, and the ICTP Trieste for hospitality and support at various stages of this work. 
C.M.~thanks the ITP Heidelberg, the Max-Planck-Institut f\"ur Gravitationsphysik, and the ICTP Trieste for hospitality.

\appendix

\section{Other ADE Singularities}\label{app:ADE}

In this appendix, we complete the discussion of the NU models describing small instantons
on ADE singularities, cf.~Table~\ref{tab:adetype}. We limit ourselves to
the $\mathrm{Spin}(32)/\ZZ_2$ heterotic string and mostly provide the resolutions
when the number $k$ of small instantons is large enough for the given patterns to be valid.
In all models the full algebra and the 
total number of blow-ups coincide with the results in Table 4 of \cite{Aspinwall:1997ye}. 
Moreover, as already explained, for a singularity of type $G$ the structure of the resolution basically follows
from the extended Dynkin diagram of $G$ \cite{Blum:1997mm}.

The results in the  $E_8 \times E_8$ case can be consulted in \cite{Font:2016odl}. It can be verified
that the quantities $h_{\mathrm R}$ and  $r_{\mathrm R}$, cf.~\eqref{checkdef} and \eqref{checkrank}, do 
match in both heterotic strings.

\subsection{\texorpdfstring{$[\mathrm{I}_{n-p-0}]$}{[I_{n-p-0}]} Model and \texorpdfstring{$\mathrm{A}_{p-1}$}{A_{p-1}} Singularities}

The number of instantons sitting on the singularity is $k=n+p$. We present the resolutions when
$n \ge n_{\rm min}$ so that $k$ is above the minimum number needed for the given result to be valid.
For $p$ even, $n_{\rm min}=p$, whereas for  $p$ odd, $n_{\rm min}=p-2$. The number of blow-ups
is $n_T=[p/2]$. The auxiliary quantity $\ell=k - 2p$ is used to simplify the displays. As shown below, for $p\ge 3$,
starting with $\mathfrak{su}(2k\text{-}8)$ the resolution includes a chain of $\mathfrak{su}$ algebras
supported at curves of self-intersection $-2$, with rank decreasing in units of eight until a final value.
For the last blow-up the self-intersection number is $-1$ and the curve supports a $\mathfrak{sp}(2\ell)$ 
or a $\mathfrak{su}(2\ell\text{+}4)$ depending on whether $p$ is even or odd. In the latter case, there is an
additional hypermultiplet in the antisymmetric representation of $\mathfrak{su}(2\ell\text{+}4)$ as required 
by anomaly cancellation \cite{Blum:1997mm}. Besides, there are hypermultiplets in $({\mathbf \fund},{\mathbf  \fund})$ 
for adjacent algebras. Notice that in all cases $h_{\mathrm R}=k p - (p^2-1)$.

%%%

\begin{equation}
\label{modA1}
\begin{tabular}{|cc|}
\multicolumn{2}{l}{\normalsize{$p=2$}}\\
\hline
$\mathfrak{sp}(k)$ &  $\mathfrak{sp}(k\text{-}4)$\\ 
1*&1\\\hline
\end{tabular}  \, .
\end{equation}

%%%

\begin{equation}
\label{modA2}
\begin{tabular}{|cc|}
\multicolumn{2}{l}{\normalsize{$p=3$}}\\
\hline
$\mathfrak{sp}(k)$ &  $\mathfrak{su}(2k\text{-}8)$\\ 
1*&1\\\hline
\end{tabular}   \, .
\end{equation}

%%%

\begin{equation}
\label{modAeven}
\begin{tabular}{|cccccc|}
\multicolumn{2}{l}{$p$ even}\\
\hline
$\mathfrak{sp}(k)$ & $\mathfrak{su}(2k\text{-}8)$ & $\mathfrak{su}(2k\text{-}16)$ &
 $\cdots$ & $\mathfrak{su}(2\ell\text{+}8)$ & $\mathfrak{sp}(\ell)$\\
1* &2&2& $\cdots$ & 2 & 1\\
 \hline
 \end{tabular}  \, .
\end{equation}

%%%
%%%

\begin{equation}
\label{modAodd}
\begin{tabular}{|cccccc|}
\multicolumn{2}{l}{$p$ odd}\\
\hline
$\mathfrak{sp}(k)$ & $\mathfrak{su}(2k\text{-}8)$ & $\mathfrak{su}(2k\text{-}16)$ &
 $\cdots$ & $\mathfrak{su}(2\ell\text{+}12)$ & $\mathfrak{su}(2\ell\text{+}4)$\\
1* &2&2& $\cdots$ & 2 & 1\\
 \hline
 \end{tabular}  \, .
\end{equation}

\subsection{\texorpdfstring{$[\mathrm{I}_n-\mathrm{I}^*_p]$}{[I_n-I^*_p]} Model and \texorpdfstring{$\mathrm{D}_{p+4}$}{D_{p+4}} Singularities}

The number of instantons at the singularity is $k=6+n+p$. We restrict to
$n \ge n_{\rm min}$ so that $k$ is large enough for the given result to be valid. This minimum value,
as well as the number of blow-ups, depends on whether $p$ is even or odd.
For $p$ even, $n_{\rm min}=p+2$, $n_T=p+4$, whereas for  $p$ odd, $n_{\rm min}=p$ and
$n_T=p+3$. For simplicity we again introduce $\ell=k - 2p$. The structure of the resolutions
is self-explanatory. In all cases $h_{\mathrm R}=k (2p + 6)  - (p+4)(2 p + 7)$.

%%%

{\footnotesize
\begin{equation}
\label{modDeven}
\begin{tabular}{|cccccccccc|}
\multicolumn{2}{l}{\normalsize{$p$ even}}\\
\hline
& $\mathfrak{sp}(k\text{-}8)$ & & &  & &  & & $\mathfrak{sp}(\ell\text{-}8)$ & \\
&1 & & &  & &  & & 1 & \\
&$\mid$ & & &  & &  & & $\mid$ & \\
 $\mathfrak{sp}(k)$ & $\mathfrak{so}(4k\text{-}16)$ & $\mathfrak{sp}(2k\text{-}16)$ & $\mathfrak{so}(4k\text{-}32)$ &
 $\mathfrak{sp}(2k\text{-}24)$ & $\cdots$ &  $\mathfrak{so}(4\ell)$ &
 $\mathfrak{sp}(2\ell\text{-}8)$ & $\mathfrak{so}(4\ell\text{-}16)$ &
$\mathfrak{sp}(\ell\text{-}8)$ \\
1*& 4 & 1 & 4 & 1 & $\cdots$ & 4 & 1&4 & 1\\
\hline
\end{tabular}\, .
\end{equation}
}

%%%
%%%

{\footnotesize
\begin{equation}
\label{modDodd}
\begin{tabular}{|ccccccccc|}
\multicolumn{4}{l}{\normalsize{$p$ odd}}\\
\hline
& $\mathfrak{sp}(k\text{-}8)$ & & &  & &  & &  \\
&1 & & &  & &  & &  \\
&$\mid$ & & &  & &  & &  \\
 $\mathfrak{sp}(k)$ & $\mathfrak{so}(4k\text{-}16)$ & $\mathfrak{sp}(2k\text{-}16)$ & $\mathfrak{so}(4k\text{-}32)$ &
 $\mathfrak{sp}(2k\text{-}24)$ & $\cdots$ &  $\mathfrak{so}(4\ell\text{-}8)$ &
 $\mathfrak{sp}(2\ell\text{-}12)$ & $\mathfrak{su}(2\ell\text{-}12)$ \\
1*& 4 & 1 & 4 & 1 & $\cdots$ & 4 & 1&2 \\
\hline
\end{tabular}\, .
\end{equation}
}

\subsection{\texorpdfstring{$[\mathrm{III}^{\ast}-\mathrm{I}_n]$}{[III^*-I_n]} Model and \texorpdfstring{$\mathrm{E}_7$}{E_7} Singularity}

The number of small instantons is $k=9+n$. The resolution for $n > 3$ has the structure of the $\mathfrak{e}_7$
extended Dynkin diagram as expected \cite{Blum:1997mm}. In this case $h_{\mathrm R}=18 k   - 133$. 
The non-generic patterns for $n \le 2$ are also shown below.

%%%

\begin{equation}
\label{modE70}
\begin{tabular}{|ccccc|}
\multicolumn{3}{l}{$n=0$}\\
\hline
$\mathfrak{sp}(9)$ & $\mathfrak{so}(20)$ & $\mathfrak{sp}(3)$ &
  $\mathfrak{so}(7)$ & $\mathfrak{su}(2)$\\
1* &4&1& 2 & 2\\
 \hline
 \end{tabular}  \, .
\end{equation}

%%%

\begin{equation}
\label{modE71}
\begin{tabular}{|cccccc|}
\multicolumn{3}{l}{$n=1$}\\
\hline
 & & & 1 &  & \\
 & & & $\mid$ &  & \\
$\mathfrak{sp}(10)$ & $\mathfrak{so}(24)$ & $\mathfrak{sp}(6)$ &
  $\mathfrak{so}(16)$ & $\mathfrak{sp}(2)$ & $\mathfrak{so}(7)$\\
 1*&4&1& 4 & 1 & 3\\
 \hline
 \end{tabular}  \, .
\end{equation}

%%%

\begin{equation}
\label{modE72}
\begin{tabular}{|cccccc|}
\multicolumn{3}{l}{$n=2$}\\
\hline
 & & & $\mathfrak{sp}(2)$ &  & \\
 & & & 1 &  & \\
 & & & $\mid$ &  & \\
$\mathfrak{sp}(11)$ & $\mathfrak{so}(28)$ & $\mathfrak{sp}(9)$ &
  $\mathfrak{so}(24)$ & $\mathfrak{sp}(5)$ & $\mathfrak{so}(12)$\\
 1*&4&1& 4 & 1 & 3\\
 \hline
 \end{tabular}  \, .
\end{equation}

%%%

\begin{equation}
\label{modE7}
\begin{tabular}{|ccccccc|}
\multicolumn{3}{l}{$n \ge 3$}\\
\hline
 & & & $\mathfrak{sp}(2k\text{-}20)$ &  &  & \\
  & & & 1 &  &  & \\
    & & & $\mid$ &  &  & \\
$\mathfrak{sp}(k)$ & $\mathfrak{so}(4k\text{-}16)$ & $\mathfrak{sp}(3k\text{-}24)$ & $\mathfrak{so}(8k\text{-}64)$ &
$\mathfrak{sp}(3k\text{-}28)$ & $\mathfrak{so}(4k\text{-}32)$ & $\mathfrak{sp}(k\text{-}12)$\\
1* &4&1& 4 & 1 & 4 & 1\\
 \hline
 \end{tabular}  \, .
\end{equation}

\bibliography{papers}
\bibliographystyle{utphys}

\end{document}